\UseRawInputEncoding
\documentclass[longauth]{aa}
\usepackage{txfonts}
\usepackage[T5,T1]{fontenc}

\usepackage[colorlinks=true]{hyperref}
\usepackage{color, soul}
\usepackage{makecell}
\hypersetup{colorlinks=true, linkcolor=blue, filecolor=blue, urlcolor=blue, citecolor=blue}

\begin{document}

\title{Search for High-Energy Neutrinos from Infrared Flares}

\author{
IceCube Collaboration:
R. Abbasi\inst{\ref{loyola}}
\and M. Ackermann\inst{\ref{zeuthen}}
\and J. Adams\inst{\ref{christchurch}}
\and J. A. Aguilar\inst{\ref{brusselslibre}}
\and M. Ahlers\inst{\ref{copenhagen}}
\and J.M. Alameddine\inst{\ref{dortmund}}
\and S. Ali\inst{\ref{kansas}}
\and N. M. Amin\inst{\ref{bartol}}
\and K. Andeen\inst{\ref{marquette}}
\and C. Arg{\"u}elles\inst{\ref{harvard}}
\and S. Athanasiadou\inst{\ref{zeuthen}}
\and S. N. Axani\inst{\ref{bartol}}
\and R. Babu\inst{\ref{michigan}}
\and X. Bai\inst{\ref{southdakota}}
\and A. Balagopal V.\inst{\ref{bartol}}
\and S. W. Barwick\inst{\ref{irvine}}
\and V. Basu\inst{\ref{utah}}
\and R. Bay\inst{\ref{berkeley}}
\and J. J. Beatty\inst{\ref{ohioastro},\ref{ohio}}
\and J. Becker Tjus\inst{\ref{bochum},\ref{a}}
\and J. Beise\inst{\ref{uppsala}}
\and C. Bellenghi\inst{\ref{munich}}
\and S. Benkel\inst{\ref{zeuthen}}
\and S. BenZvi\inst{\ref{rochester}}
\and D. Berley\inst{\ref{maryland}}
\and E. Bernardini\inst{\ref{padova},\ref{b}}
\and D. Z. Besson\inst{\ref{kansas}}
\and E. Blaufuss\inst{\ref{maryland}}
\and L. Bloom\inst{\ref{alabama}}
\and S. Blot\inst{\ref{zeuthen}}
\and F. Bontempo\inst{\ref{karlsruhe}}
\and J. Y. Book Motzkin\inst{\ref{harvard}}
\and C. Boscolo Meneguolo\inst{\ref{padova},\ref{b}}
\and S. B{\"o}ser\inst{\ref{mainz}}
\and O. Botner\inst{\ref{uppsala}}
\and J. Braun\inst{\ref{madisonpac}}
\and B. Brinson\inst{\ref{maryland}}
\and Z. Brisson-Tsavoussis\inst{\ref{queens}}
\and L. Brusa\inst{\ref{erlangen}}
\and R. T. Burley\inst{\ref{adelaide}}
\and D. Butterfield\inst{\ref{madisonpac}}
\and K. Carloni\inst{\ref{harvard}}
\and J. Carpio\inst{\ref{copenhagen},\ref{lasvegasphysics},\ref{lasvegasastro}}
\and N. Chau\inst{\ref{brusselslibre}}
\and Y. C. Chen\inst{\ref{bartol}}
\and Z. Chen\inst{\ref{stonybrook}}
\and D. Chirkin\inst{\ref{madisonpac}}
\and S. Choi\inst{\ref{utah}}
\and A. Chubarov\inst{\ref{erlangen}}
\and B. A. Clark\inst{\ref{maryland}}
\and D. A. Coloma Borja\inst{\ref{padova}}
\and A. Connolly\inst{\ref{ohioastro},\ref{ohio}}
\and J. M. Conrad\inst{\ref{mit}}
\and D. F. Cowen\inst{\ref{pennastro},\ref{pennphys}}
\and C. De Clercq\inst{\ref{brusselsvrije}}
\and J. J. DeLaunay\inst{\ref{pennastro}}
\and D. Delgado\inst{\ref{harvard}}
\and T. Delmeulle\inst{\ref{brusselslibre}}
\and S. Deng\inst{\ref{aachen}}
\and P. Desiati\inst{\ref{madisonpac}}
\and K. D. de Vries\inst{\ref{brusselsvrije}}
\and G. de Wasseige\inst{\ref{uclouvain-2026}}
\and T. DeYoung\inst{\ref{michigan}}
\and J. C. D{\'\i}az-V{\'e}lez\inst{\ref{madisonpac}}
\and S. DiKerby\inst{\ref{michigan}}
\and T. Ding\inst{\ref{lasvegasphysics},\ref{lasvegasastro}}
\and M. Dittmer\inst{\ref{munster-2024}}
\and A. Domi\inst{\ref{erlangen}}
\and L. Draper\inst{\ref{utah}}
\and L. Dueser\inst{\ref{aachen}}
\and D. Durnford\inst{\ref{edmonton}}
\and K. Dutta\inst{\ref{mainz}}
\and M. A. DuVernois\inst{\ref{madisonpac}}
\and T. Ehrhardt\inst{\ref{mainz}}
\and L. Eidenschink\inst{\ref{munich}}
\and A. Eimer\inst{\ref{erlangen}}
\and C. Eldridge\inst{\ref{gent}}
\and P. Eller\inst{\ref{munich}}
\and E. Ellinger\inst{\ref{wuppertal}}
\and D. Els{\"a}sser\inst{\ref{dortmund}}
\and R. Engel\inst{\ref{karlsruhe},\ref{karlsruheexp}}
\and H. Erpenbeck\inst{\ref{madisonpac}}
\and W. Esmail\inst{\ref{munster-2024}}
\and S. Eulig\inst{\ref{harvard}}
\and J. Evans\inst{\ref{maryland}}
\and P. A. Evenson\inst{\ref{bartol}}
\and K. L. Fan\inst{\ref{maryland}}
\and K. Fang\inst{\ref{madisonpac}}
\and K. Farrag\inst{\ref{chiba2022}}
\and A. Fattorini\inst{\ref{dortmund}}
\and A. R. Fazely\inst{\ref{southern}}
\and A. Fedynitch\inst{\ref{sinica}}
\and N. Feigl\inst{\ref{berlin}}
\and C. Finley\inst{\ref{stockholmokc}}
\and D. Fox\inst{\ref{pennastro}}
\and A. Franckowiak\inst{\ref{bochum}}
\and S. Fukami\inst{\ref{zeuthen}}
\and P. F{\"u}rst\inst{\ref{aachen}}
\and J. Gallagher\inst{\ref{madisonastro}}
\and A. Garcia\inst{\ref{harvard}}
\and M. Garcia\inst{\ref{bartol}}
\and E. Genton\inst{\ref{brusselslibre},\ref{harvard}}
\and L. Gerhardt\inst{\ref{lbnl}}
\and A. Ghadimi\inst{\ref{alabama}}
\and C. Glaser\inst{\ref{dortmund},\ref{uppsala}}
\and T. Gl{\"u}senkamp\inst{\ref{stockholmokc}}
\and J. G. Gonzalez\inst{\ref{bartol}}
\and S. Goswami\inst{\ref{lasvegasphysics},\ref{lasvegasastro}}
\and A. Granados\inst{\ref{michigan}}
\and D. Grant\inst{\ref{simon-fraser-2024-2}}
\and S. J. Gray\inst{\ref{maryland}}
\and S. Griffin\inst{\ref{madisonpac}}
\and S. Griswold\inst{\ref{madisonpac}}
\and K. M. Groth\inst{\ref{copenhagen}}
\and D. Guevel\inst{\ref{madisonpac}}
\and C. G{\"u}nther\inst{\ref{aachen}}
\and P. Gutjahr\inst{\ref{dortmund}}
\and C. Ha\inst{\ref{chung-ang-2024}}
\and A. Hallgren\inst{\ref{uppsala}}
\and F. Halzen\inst{\ref{madisonpac}}
\and M. Handt\inst{\ref{aachen}}
\and K. Hanson\inst{\ref{madisonpac}}
\and J. Hardin\inst{\ref{mit}}
\and A. A. Harnisch\inst{\ref{michigan}}
\and P. Hatch\inst{\ref{queens}}
\and A. Haungs\inst{\ref{karlsruhe}}
\and J. H{\"a}u{\ss}ler\inst{\ref{aachen}}
\and K. Helbing\inst{\ref{wuppertal}}
\and J. Hellrung\inst{\ref{bochum}}
\and B. Henke\inst{\ref{michigan}}
\and L. Hennig\inst{\ref{erlangen}}
\and F. Henningsen\inst{\ref{erlangen}}
\and L. Heuermann\inst{\ref{aachen}}
\and R. Hewett\inst{\ref{christchurch}}
\and N. Heyer\inst{\ref{uppsala}}
\and S. Hickford\inst{\ref{wuppertal}}
\and A. Hidvegi\inst{\ref{stockholmokc}}
\and C. Hill\inst{\ref{munich}}
\and G. C. Hill\inst{\ref{adelaide}}
\and R. Hmaid\inst{\ref{chiba2022}}
\and K. D. Hoffman\inst{\ref{maryland}}
\and A. Hollnagel\inst{\ref{chiba2022}}
\and D. Hooper\inst{\ref{madisonpac}}
\and S. Hori\inst{\ref{madisonpac}}
\and K. Hoshina\inst{\ref{madisonpac},\ref{c}}
\and M. Hostert\inst{\ref{harvard}}
\and W. Hou\inst{\ref{karlsruhe}}
\and M. Hrywniak\inst{\ref{stockholmokc}}
\and T. Huber\inst{\ref{karlsruhe}}
\and K. Hultqvist\inst{\ref{stockholmokc}}
\and K. Hymon\inst{\ref{sinica}}
\and A. Ishihara\inst{\ref{chiba2022}}
\and W. Iwakiri\inst{\ref{chiba2022}}
\and M. Jacquart\inst{\ref{copenhagen}}
\and S. Jain\inst{\ref{madisonpac}}
\and O. Janik\inst{\ref{erlangen}}
\and M. Jansson\inst{\ref{uclouvain-2026}}
\and M. Jin\inst{\ref{harvard}}
\and N. Kamp\inst{\ref{harvard}}
\and D. Kang\inst{\ref{karlsruhe}}
\and W. Kang\inst{\ref{drexel}}
\and A. Kappes\inst{\ref{munster-2024}}
\and L. Kardum\inst{\ref{dortmund}}
\and T. Karg\inst{\ref{zeuthen}}
\and A. Karle\inst{\ref{madisonpac}}
\and A. Katil\inst{\ref{edmonton}}
\and M. Kauer\inst{\ref{madisonpac}}
\and J. L. Kelley\inst{\ref{madisonpac}}
\and B. Kern\inst{\ref{aachen}}
\and M. Khanal\inst{\ref{utah}}
\and A. Khatee Zathul\inst{\ref{madisonpac}}
\and A. Kheirandish\inst{\ref{lasvegasphysics},\ref{lasvegasastro}}
\and T. Kim\inst{\ref{skku}}
\and H. Kimku\inst{\ref{chung-ang-2024}}
\and F. Kirchner\inst{\ref{erlangen}}
\and J. Kiryluk\inst{\ref{stonybrook}}
\and C. Klein\inst{\ref{zeuthen}}
\and S. R. Klein\inst{\ref{berkeley},\ref{lbnl}}
\and Y. Kobayashi\inst{\ref{chiba2022}}
\and S. Koch\inst{\ref{erlangen}}
\and A. Kochocki\inst{\ref{michigan}}
\and R. Koirala\inst{\ref{bartol}}
\and H. Kolanoski\inst{\ref{berlin}}
\and T. Kontrimas\inst{\ref{munich}}
\and L. K{\"o}pke\inst{\ref{mainz}}
\and C. Kopper\inst{\ref{erlangen}}
\and D. J. Koskinen\inst{\ref{copenhagen}}
\and P. Koundal\inst{\ref{bartol}}
\and M. Kowalski\inst{\ref{berlin},\ref{zeuthen}}
\and A. Kravka\inst{\ref{utah}}
\and N. Krieger\inst{\ref{bochum}}
\and T. Krishnan\inst{\ref{harvard}}
\and K. Kruiswijk\inst{\ref{uclouvain-2026}}
\and E. Krupczak\inst{\ref{michigan}}
\and E. Kun\inst{\ref{bochum}}
\and N. Kurahashi\inst{\ref{drexel}}
\and C. Lagunas Gualda\inst{\ref{erlangen}}
\and L. Lallement Arnaud\inst{\ref{brusselslibre}}
\and M. J. Larson\inst{\ref{maryland}}
\and F. Lauber\inst{\ref{wuppertal}}
\and J. P. Lazar\inst{\ref{uclouvain-2026}}
\and K. Leonard DeHolton\inst{\ref{pennphys}}
\and A. Leszczy{\'n}ska\inst{\ref{bartol}}
\and C. Li\inst{\ref{madisonpac}}
\and J. Liao\inst{\ref{georgia}}
\and C. Lin\inst{\ref{bartol}}
\and Q. R. Liu\inst{\ref{simon-fraser-2024-2}}
\and Y. T. Liu\inst{\ref{pennphys}}
\and M. Liubarska\inst{\ref{edmonton}}
\and C. Love\inst{\ref{drexel}}
\and L. Lu\inst{\ref{madisonpac}}
\and F. Lucarelli\inst{\ref{geneva}}
\and W. Luszczak\inst{\ref{ohioastro},\ref{ohio}}
\and Y. Lyu\inst{\ref{berkeley},\ref{lbnl}}
\and M. Macdonald\inst{\ref{harvard}}
\and F. Machingal\inst{\ref{southdakota}}
\and E. Magnus\inst{\ref{brusselsvrije}}
\and Y. Makino\inst{\ref{madisonpac}}
\and E. Manao\inst{\ref{munich}}
\and S. Mancina\inst{\ref{padova},\ref{d}}
\and A. Mand\inst{\ref{madisonpac}}
\and I. C. Mari{\c{s}}\inst{\ref{brusselslibre}}
\and S. Marka\inst{\ref{columbia}}
\and Z. Marka\inst{\ref{columbia}}
\and I. Martinez-Soler\inst{\ref{harvard}}
\and R. Maruyama\inst{\ref{yale}}
\and J. Mauro\inst{\ref{uclouvain-2026}}
\and F. Mayhew\inst{\ref{michigan}}
\and F. McNally\inst{\ref{mercer}}
\and K. Meagher\inst{\ref{madisonpac}}
\and A. Medina\inst{\ref{ohio}}
\and M. Meier\inst{\ref{chiba2022}}
\and Y. Merckx\inst{\ref{brusselsvrije}}
\and L. Merten\inst{\ref{bochum}}
\and J. M. Mitchell\inst{\ref{southern}}
\and S. Mondal\inst{\ref{utah}}
\and T. Montaruli\inst{\ref{geneva}}
\and R. W. Moore\inst{\ref{edmonton}}
\and Y. Morii\inst{\ref{chiba2022}}
\and A. Mosbrugger\inst{\ref{erlangen}}
\and D. Mousadi\inst{\ref{zeuthen}}
\and E. Moyaux\inst{\ref{uclouvain-2026}}
\and T. Mukherjee\inst{\ref{karlsruhe}}
\and M. Nakos\inst{\ref{madisonpac}}
\and U. Naumann\inst{\ref{wuppertal}}
\and R. Neshat\inst{\ref{utah}}
\and L. Neste\inst{\ref{stockholmokc}}
\and M. Neumann\inst{\ref{munster-2024}}
\and M. U. Nisa\inst{\ref{michigan}}
\and K. Noda\inst{\ref{chiba2022}}
\and A. Noell\inst{\ref{aachen}}
\and A. Novikov\inst{\ref{bartol}}
\and A. Obertacke\inst{\ref{stockholmokc}}
\and V. O'Dell\inst{\ref{madisonpac}}
\and A. Olivas\inst{\ref{maryland}}
\and R. Orsoe\inst{\ref{munich}}
\and J. Osborn\inst{\ref{madisonpac}}
\and E. O'Sullivan\inst{\ref{uppsala}}
\and B. Owens\inst{\ref{queens}}
\and V. Palusova\inst{\ref{mainz}}
\and H. Pandya\inst{\ref{bartol}}
\and A. Parenti\inst{\ref{brusselslibre}}
\and C. Parisel\inst{\ref{madisonpac}}
\and N. Park\inst{\ref{queens}}
\and V. Parrish\inst{\ref{michigan}}
\and E. N. Paudel\inst{\ref{alabama}}
\and L. Paul\inst{\ref{southdakota}}
\and C. P{\'e}rez de los Heros\inst{\ref{uppsala}}
\and T. Pernice\inst{\ref{zeuthen}}
\and T. C. Petersen\inst{\ref{copenhagen}}
\and J. Peterson\inst{\ref{madisonpac}}
\and S. Pick\inst{\ref{zeuthen}}
\and M. Plum\inst{\ref{southdakota}}
\and A. Pont{\'e}n\inst{\ref{uppsala}}
\and V. Poojyam\inst{\ref{alabama}}
\and B. Pries\inst{\ref{michigan}}
\and R. Procter-Murphy\inst{\ref{maryland}}
\and G. T. Przybylski\inst{\ref{lbnl}}
\and L. Pyras\inst{\ref{utah}}
\and C. Raab\inst{\ref{uclouvain-2026}}
\and J. Rack-Helleis\inst{\ref{mainz}}
\and N. Rad\inst{\ref{zeuthen}}
\and M. Ravn\inst{\ref{uppsala}}
\and K. Rawlins\inst{\ref{anchorage}}
\and Z. Rechav\inst{\ref{madisonpac}}
\and A. Rehman\inst{\ref{bartol}}
\and R. Reimann\inst{\ref{aachen}}
\and I. Reistroffer\inst{\ref{southdakota}}
\and E. Resconi\inst{\ref{munich}}
\and C. D. Rho\inst{\ref{skku}}
\and W. Rhode\inst{\ref{dortmund}}
\and L. Ricca\inst{\ref{uclouvain-2026}}
\and B. Riedel\inst{\ref{madisonpac}}
\and A. Rifaie\inst{\ref{wuppertal}}
\and E. J. Roberts\inst{\ref{adelaide}}
\and S. Rodan\inst{\ref{riverfalls}}
\and M. Rongen\inst{\ref{erlangen}}
\and A. Rosted\inst{\ref{chiba2022}}
\and C. Rott\inst{\ref{utah}}
\and T. Ruhe\inst{\ref{dortmund}}
\and L. Ruohan\inst{\ref{munich}}
\and D. Ryckbosch\inst{\ref{gent}}
\and J. Saffer\inst{\ref{karlsruheexp}}
\and D. Salazar-Gallegos\inst{\ref{michigan}}
\and P. Sampathkumar\inst{\ref{karlsruhe}}
\and A. Sandrock\inst{\ref{wuppertal}}
\and G. Sanger-Johnson\inst{\ref{michigan}}
\and M. Santander\inst{\ref{alabama}}
\and S. Sarkar\inst{\ref{oxford}}
\and M. Scarnera\inst{\ref{uclouvain-2026}}
\and M. Schaufel\inst{\ref{aachen}}
\and H. Schieler\inst{\ref{karlsruhe}}
\and S. Schindler\inst{\ref{erlangen}}
\and L. Schlickmann\inst{\ref{mainz}}
\and B. Schl{\"u}ter\inst{\ref{munster-2024}}
\and F. Schl{\"u}ter\inst{\ref{brusselslibre}}
\and N. Schmeisser\inst{\ref{wuppertal}}
\and T. Schmidt\inst{\ref{maryland}}
\and F. Schmitt\inst{\ref{karlsruheexp}}
\and A. Scholz\inst{\ref{munich}}
\and F. G. Schr{\"o}der\inst{\ref{karlsruhe},\ref{bartol}}
\and S. Schwirn\inst{\ref{aachen}}
\and S. Sclafani\inst{\ref{maryland}}
\and D. Seckel\inst{\ref{bartol}}
\and L. Seen\inst{\ref{madisonpac}}
\and M. Seikh\inst{\ref{kansas}}
\and S. Seunarine\inst{\ref{riverfalls}}
\and P. A. Sevle Myhr\inst{\ref{uclouvain-2026}}
\and R. Shah\inst{\ref{drexel}}
\and S. Shah\inst{\ref{rochester}}
\and A. Sharma\inst{\ref{christchurch}}
\and S. Shefali\inst{\ref{harvard}}
\and N. Shimizu\inst{\ref{chiba2022}}
\and M. Shin\inst{\ref{skku}}
\and R. Singh\inst{\ref{aachen}}
\and B. Skrzypek\inst{\ref{berkeley}}
\and R. Snihur\inst{\ref{madisonpac}}
\and J. Soedingrekso\inst{\ref{dortmund}}
\and D. Soldin\inst{\ref{utah}}
\and P. Soldin\inst{\ref{aachen}}
\and G. Sommani\inst{\ref{bochum}}
\and D. Song\inst{\ref{brusselslibre}}
\and C. Spannfellner\inst{\ref{munich}}
\and G. M. Spiczak\inst{\ref{riverfalls}}
\and C. Spiering\inst{\ref{zeuthen}}
\and J. Stachurska\inst{\ref{gent}}
\and M. Stamatikos\inst{\ref{ohio}}
\and T. Stanev\inst{\ref{bartol}}
\and T. Stezelberger\inst{\ref{lbnl}}
\and T. St{\"u}rwald\inst{\ref{wuppertal}}
\and T. Stuttard\inst{\ref{copenhagen}}
\and G. W. Sullivan\inst{\ref{maryland}}
\and I. Taboada\inst{\ref{georgia}}
\and S. Ter-Antonyan\inst{\ref{southern}}
\and A. Terliuk\inst{\ref{munich}}
\and A. Thakuri\inst{\ref{southdakota}}
\and P. Thelen\inst{\ref{aachen}}
\and M. Thiesmeyer\inst{\ref{madisonpac}}
\and W. G. Thompson\inst{\ref{harvard}}
\and J. Thwaites\inst{\ref{queens}}
\and W. Tian\inst{\ref{madisonpac}}
\and S. Tilav\inst{\ref{bartol}}
\and K. Tollefson\inst{\ref{michigan}}
\and J. A. Torres\inst{\ref{utah}}
\and S. Toscano\inst{\ref{brusselslibre}}
\and D. Tosi\inst{\ref{madisonpac}}
\and K. Upshaw\inst{\ref{southern}}
\and A. Vaidyanathan\inst{\ref{marquette}}
\and N. Valtonen-Mattila\inst{\ref{bochum}}
\and J. Valverde\inst{\ref{marquette}}
\and J. Vandenbroucke\inst{\ref{madisonpac}}
\and T. Van Eeden\inst{\ref{zeuthen}}
\and N. van Eijndhoven\inst{\ref{brusselsvrije}}
\and L. Van Rootselaar\inst{\ref{dortmund}}
\and J. van Santen\inst{\ref{zeuthen}}
\and J. Vara\inst{\ref{munster-2024}}
\and F. Varsi\inst{\ref{karlsruheexp}}
\and M. Velazquez\inst{\ref{georgia}}
\and M. Venugopal\inst{\ref{karlsruhe}}
\and M. Vereecken\inst{\ref{gent}}
\and S. Vergara Carrasco\inst{\ref{christchurch}}
\and S. Verpoest\inst{\ref{bartol}}
\and A. Vijai\inst{\ref{maryland}}
\and J. Villarreal\inst{\ref{mit}}
\and C. Walck\inst{\ref{stockholmokc}}
\and A. Wang\inst{\ref{georgia}}
\and E. H. S. Warrick\inst{\ref{alabama}}
\and C. Weaver\inst{\ref{michigan}}
\and P. Weigel\inst{\ref{harvard}}
\and A. Weindl\inst{\ref{karlsruhe}}
\and J. Weldert\inst{\ref{mainz}}
\and A. Y. Wen\inst{\ref{harvard}}
\and C. Wendt\inst{\ref{madisonpac}}
\and J. Werthebach\inst{\ref{dortmund}}
\and M. Weyrauch\inst{\ref{karlsruhe}}
\and N. Whitehorn\inst{\ref{michigan}}
\and C. H. Wiebusch\inst{\ref{aachen}}
\and D. R. Williams\inst{\ref{alabama}}
\and L. Witthaus\inst{\ref{dortmund}}
\and J. Woodward\inst{\ref{mit}}
\and G. Wrede\inst{\ref{erlangen}}
\and X. W. Xu\inst{\ref{southern}}
\and J. P. Yanez\inst{\ref{edmonton}}
\and Y. Yao\inst{\ref{madisonpac}}
\and E. Yildizci\inst{\ref{madisonpac}}
\and S. Yoshida\inst{\ref{chiba2022}}
\and F. Yu\inst{\ref{harvard}}
\and S. Yu\inst{\ref{utah}}
\and T. Yuan\inst{\ref{madisonpac}}
\and S. Yun-C{\'a}rcamo\inst{\ref{drexel}}
\and A. Zander Jurowitzki\inst{\ref{munich}}
\and A. Zegarelli\inst{\ref{bochum}}
\and S. Zhang\inst{\ref{michigan}}
\and Z. Zhang\inst{\ref{stonybrook}}
\and P. Zhelnin\inst{\ref{harvard}}
\and P. Zilberman\inst{\ref{madisonpac}}
\and C. Zilleruelo Ca{\~n}as\inst{\ref{zeuthen}}
\and M. Zimmermann\inst{\ref{aachen}}
}
\institute{III. Physikalisches Institut, RWTH Aachen University, D-52056 Aachen, Germany \label{aachen} 
\and Department of Physics, University of Adelaide, Adelaide, 5005, Australia \label{adelaide} 
\and Dept. of Physics and Astronomy, University of Alaska Anchorage, 3211 Providence Dr., Anchorage, AK 99508, USA \label{anchorage} 
\and School of Physics and Center for Relativistic Astrophysics, Georgia Institute of Technology, Atlanta, GA 30332, USA \label{georgia} 
\and Dept. of Physics, Southern University, Baton Rouge, LA 70813, USA \label{southern} 
\and Dept. of Physics, University of California, Berkeley, CA 94720, USA \label{berkeley} 
\and Lawrence Berkeley National Laboratory, Berkeley, CA 94720, USA \label{lbnl} 
\and Institut f{\"u}r Physik, Humboldt-Universit{\"a}t zu Berlin, D-12489 Berlin, Germany \label{berlin} 
\and Fakult{\"a}t f{\"u}r Physik {\&} Astronomie, Ruhr-Universit{\"a}t Bochum, D-44780 Bochum, Germany \label{bochum} 
\and Universit{\'e} Libre de Bruxelles, Science Faculty CP230, B-1050 Brussels, Belgium \label{brusselslibre} 
\and Vrije Universiteit Brussel (VUB), Dienst ELEM, B-1050 Brussels, Belgium \label{brusselsvrije} 
\and Dept. of Physics, Simon Fraser University, Burnaby, BC V5A 1S6, Canada \label{simon-fraser-2024-2} 
\and Department of Physics and Laboratory for Particle Physics and Cosmology, Harvard University, Cambridge, MA 02138, USA \label{harvard} 
\and Dept. of Physics, Massachusetts Institute of Technology, Cambridge, MA 02139, USA \label{mit} 
\and Dept. of Physics and The International Center for Hadron Astrophysics, Chiba University, Chiba 263-8522, Japan \label{chiba2022} 
\and Department of Physics, Loyola University Chicago, Chicago, IL 60660, USA \label{loyola} 
\and Dept. of Physics and Astronomy, University of Canterbury, Private Bag 4800, Christchurch, New Zealand \label{christchurch} 
\and Dept. of Physics, University of Maryland, College Park, MD 20742, USA \label{maryland} 
\and Dept. of Astronomy, Ohio State University, Columbus, OH 43210, USA \label{ohioastro} 
\and Dept. of Physics and Center for Cosmology and Astro-Particle Physics, Ohio State University, Columbus, OH 43210, USA \label{ohio} 
\and Niels Bohr Institute, University of Copenhagen, DK-2100 Copenhagen, Denmark \label{copenhagen} 
\and Dept. of Physics, TU Dortmund University, D-44221 Dortmund, Germany \label{dortmund} 
\and Dept. of Physics and Astronomy, Michigan State University, East Lansing, MI 48824, USA \label{michigan} 
\and Dept. of Physics, University of Alberta, Edmonton, Alberta, T6G 2E1, Canada \label{edmonton} 
\and Erlangen Centre for Astroparticle Physics, Friedrich-Alexander-Universit{\"a}t Erlangen-N{\"u}rnberg, D-91058 Erlangen, Germany \label{erlangen} 
\and Physik-department, Technische Universit{\"a}t M{\"u}nchen, D-85748 Garching, Germany \label{munich} 
\and D{\'e}partement de physique nucl{\'e}aire et corpusculaire, Universit{\'e} de Gen{\`e}ve, CH-1211 Gen{\`e}ve, Switzerland \label{geneva} 
\and Dept. of Physics and Astronomy, University of Gent, B-9000 Gent, Belgium \label{gent} 
\and Dept. of Physics and Astronomy, University of California, Irvine, CA 92697, USA \label{irvine} 
\and Karlsruhe Institute of Technology, Institute for Astroparticle Physics, D-76021 Karlsruhe, Germany \label{karlsruhe} 
\and Karlsruhe Institute of Technology, Institute of Experimental Particle Physics, D-76021 Karlsruhe, Germany \label{karlsruheexp} 
\and Dept. of Physics, Engineering Physics, and Astronomy, Queen's University, Kingston, ON K7L 3N6, Canada \label{queens} 
\and Department of Physics {\&} Astronomy, University of Nevada, Las Vegas, NV 89154, USA \label{lasvegasphysics} 
\and Nevada Center for Astrophysics, University of Nevada, Las Vegas, NV 89154, USA \label{lasvegasastro} 
\and Dept. of Physics and Astronomy, University of Kansas, Lawrence, KS 66045, USA \label{kansas} 
\and UCLouvain, Centre for Cosmology, Particle Physics and Phenomenology, CP3, Chemin du Cyclotron 2, 1348 Louvain-la-Neuve, Belgium \label{uclouvain-2026} 
\and Department of Physics, Mercer University, Macon, GA 31207-0001, USA \label{mercer} 
\and Dept. of Astronomy, University of Wisconsin{\textemdash}Madison, Madison, WI 53706, USA \label{madisonastro} 
\and Dept. of Physics and Wisconsin IceCube Particle Astrophysics Center, University of Wisconsin{\textemdash}Madison, Madison, WI 53706, USA \label{madisonpac} 
\and Institute of Physics, University of Mainz, Staudinger Weg 7, D-55099 Mainz, Germany \label{mainz} 
\and Department of Physics, Marquette University, Milwaukee, WI 53201, USA \label{marquette} 
\and Institut f{\"u}r Kernphysik, Universit{\"a}t M{\"u}nster, D-48149 M{\"u}nster, Germany \label{munster-2024} 
\and Bartol Research Institute and Dept. of Physics and Astronomy, University of Delaware, Newark, DE 19716, USA \label{bartol} 
\and Dept. of Physics, Yale University, New Haven, CT 06520, USA \label{yale} 
\and Columbia Astrophysics and Nevis Laboratories, Columbia University, New York, NY 10027, USA \label{columbia} 
\and Dept. of Physics, University of Oxford, Parks Road, Oxford OX1 3PU, United Kingdom \label{oxford} 
\and Dipartimento di Fisica e Astronomia Galileo Galilei, Universit{\`a} Degli Studi di Padova, I-35122 Padova PD, Italy \label{padova} 
\and Dept. of Physics, Drexel University, 3141 Chestnut Street, Philadelphia, PA 19104, USA \label{drexel} 
\and Physics Department, South Dakota School of Mines and Technology, Rapid City, SD 57701, USA \label{southdakota} 
\and Dept. of Physics, University of Wisconsin, River Falls, WI 54022, USA \label{riverfalls} 
\and Dept. of Physics and Astronomy, University of Rochester, Rochester, NY 14627, USA \label{rochester} 
\and Department of Physics and Astronomy, University of Utah, Salt Lake City, UT 84112, USA \label{utah} 
\and Dept. of Physics, Chung-Ang University, Seoul 06974, Republic of Korea \label{chung-ang-2024} 
\and Oskar Klein Centre and Dept. of Physics, Stockholm University, SE-10691 Stockholm, Sweden \label{stockholmokc} 
\and Dept. of Physics and Astronomy, Stony Brook University, Stony Brook, NY 11794-3800, USA \label{stonybrook} 
\and Dept. of Physics, Sungkyunkwan University, Suwon 16419, Republic of Korea \label{skku} 
\and Institute of Physics, Academia Sinica, Taipei, 11529, Taiwan \label{sinica} 
\and Dept. of Physics and Astronomy, University of Alabama, Tuscaloosa, AL 35487, USA \label{alabama} 
\and Dept. of Astronomy and Astrophysics, Pennsylvania State University, University Park, PA 16802, USA \label{pennastro} 
\and Dept. of Physics, Pennsylvania State University, University Park, PA 16802, USA \label{pennphys} 
\and Dept. of Physics and Astronomy, Uppsala University, Box 516, SE-75120 Uppsala, Sweden \label{uppsala} 
\and Dept. of Physics, University of Wuppertal, D-42119 Wuppertal, Germany \label{wuppertal} 
\and Deutsches Elektronen-Synchrotron DESY, Platanenallee 6, D-15738 Zeuthen, Germany \label{zeuthen} 
\and also at Department of Space, Earth and Environment, Chalmers University of Technology, 412 96 Gothenburg, Sweden\label{a} 
\and also at INFN Padova, I-35131 Padova, Italy\label{b} 
\and also at Earthquake Research Institute, University of Tokyo, Bunkyo, Tokyo 113-0032, Japan\label{c} 
\and now at INFN Padova, I-35131 Padova, Italy\label{d} 
}

\date{Received ; accepted }
   
\abstract
   {IceCube has detected a diffuse flux of high-energy astrophysical neutrinos. While its astrophysical origin is unknown, a significant neutrino emission has been observed from the gamma-ray blazar TXS~0506+056 and the Seyfert galaxy NGC~1068. This suggests accretion by supermassive black holes as one of the candidate mechanisms of the diffuse neutrino flux. Indeed, external studies that made use of IceCube data identified candidate sources, including AT2019dsg, likely a stellar tidal disruption event (TDE), and AT2019fdr, associated with an active galactic nucleus (AGN). An additional study using an optically detected sample of 63 accretion flares revealed another candidate as a potential high-energy neutrino counterpart: AT2019aalc. All of these candidate transients exhibit strong dust echoes, characterized by delayed infrared (IR) emission relative to the initial optical outburst.
     }
   {We look for correlations between the most comprehensive catalog of infrared flares and 10 years of IceCube archival data.}
   {We perform a time-dependent stacking analysis testing 393 sources in the Northern Sky against 10 years of through-going tracks.}
   {No significant neutrino emission is observed, and we derive upper limits to constrain neutrino emission from this catalog. These limits are subsequently extrapolated to a population-level study.}
   {We find that the contribution of IR flares to the IceCube diffuse neutrino flux at 100 TeV is below 2.7$\%$ for a spectral index of 2.5.}


\maketitle

%________________________________________________________________

\section{Introduction}
\label{sec:introduction}

The IceCube Neutrino Observatory is a neutrino detector located near the Amundsen-Scott South Pole station. It consists of 5160 Digital Optical Modules (DOMs) attached to 86 vertical support and read-out cables (called "strings") that instrument one cubic kilometer of glacial ice at a depth of 1,500 to 2,500 meters below the surface \citep{Aartsen_Ackermann_Adams_Aguilar_Ahlers_Ahrens_Altmann_Andeen_Anderson_Ansseau_et_al._2017}. These detection units record Cherenkov radiation emitted by charged secondary particles produced in the interactions between neutrinos and the surrounding ice. During the course of its operation, IceCube has measured a diffuse flux of high-energy astrophysical neutrinos whose sources remain largely unidentified \citep{IceCube_Collaboration_2013}. To date, the most significant identifications of extragalactic high-energy neutrino sources are associated with accreting supermassive black holes (SMBHs). These include the flaring blazar TXS~0506+056, identified through multi-messenger coincidence with a high-energy neutrino alert and further supported by an independent archival excess found in prior IceCube data \citep{TXS2018, IceCube_Collaboration_Aartsen_Ackermann_Adams_Aguilar_Ahlers_Ahrens_Al_Samarai_Altmann_Andeen_et_al._2018}, and the nearby Seyfert galaxy NGC~1068, observed as a persistent neutrino excess in time-integrated IceCube searches \citep{icecube2022evidence, Abbasi_Ackermann_Adams_Agarwalla_Aguilar_Ahlers_Alameddine_Ali_Amin_Andeen_et_al._2026}. In addition, IceCube has recently reported the first observation of high-energy neutrino emission from the Galactic Plane \citep{icecube2023observation}. \\

Tidal disruption events (TDEs) have been proposed as potential sources of high-energy astrophysical neutrinos \citep{farrar2009giant, Wang_Liu_Dai_Cheng_2011}. TDEs occur when a star passes sufficiently close to a SMBH and is torn apart by the black hole gravitational tidal forces. Following this process, approximately half of the stellar debris remains bound to, and eventually falls back toward, the SMBH, forming a transient accretion flow \citep{Rees_1988, Evans_Kochanek_1989}. This process powers a luminous electromagnetic flare observed across optical, ultraviolet (UV) and X-ray wavelengths, over timescales of months to years \citep{Komossa_Greiner_1999, Gezari_2006, Gezari_2021}. In some cases, TDEs also exhibit radio emission associated with relativistic jets, which have been proposed as promising sites for cosmic-ray acceleration and high-energy neutrino production \citep{Bloom_Giannios_Metzger_Cenko_Perley_Butler_Tanvir_Levan_Brien_Strubbe_et_al._2011, Burrows_Kennea_Ghisellini_Mangano_Zhang_Page_Eracleous_Romano_Sakamoto_Falcone_et_al._2011}. In addition to the primary emission, TDEs can produce infrared (IR) flares through dust reprocessing: radiation from the accretion phase is absorbed by dust in the vicinity of the SMBH and re-emitted thermally in the IR, giving rise to a dust-echo \citep{Lu_Kumar_Evans_2016, van_Velzen_Mendez_Krolik_Gorjian_2016, Jiang_Dou_Wang_Yang_Lyu_Zhou_2016}. Theoretical models have proposed that IR dust echoes may provide the target photon field for high-energy cosmic-ray interactions leading to delayed neutrino emission, complementing that concurrent with the X-ray emission from the accretion disk and corona or optical-UV (OUV) radiation from the photosphere \citep{Winter_Lunardini_2023}. 

To investigate neutrino emission from transient phenomena, IceCube has operated a real-time alert system for more than a decade \citep{Aartsen:2016lmt}, issuing notifications for track-like events with a high probability of astrophysical origin. Follow-up observations of these alerts with the Zwicky Transient Facility (ZTF; \citealt{bellm2019zwicky}) and dedicated counterpart-search programs have enabled systematic searches from transient electromagnetic counterparts \citep{Necker_de_Jaeger_Stein_Franckowiak_Shappee_Kowalski_Kochanek_Stanek_Beacom_Desai_et_al._2022, Stein_Reusch_Franckowiak_Kowalski_Necker_Weimann_Kasliwal_Sollerman_Ahumada_Amaro_Seoane_et_al._2023}. External studies that made use of IceCube data found two IceCube high-energy neutrino alerts in coincidence with events exhibiting time variable emission: the TDE AT2019dsg \citep{Stein_Velzen_Kowalski_Franckowiak_Gezari_Miller-Jones_Frederick_Sfaradi_Bietenholz_Horesh_et_al} and AT2019fdr \citep{reusch2022candidate}, both associated with active galactic nuclei (AGNs). Both events exhibit delayed IR emission relative to the optical signal. An additional study using a sample of 63 accretion flares identified with ZTF revealed another candidate high-energy neutrino counterpart, AT2019aalc, a giant AGN flare. Similar to AT2019dsg and AT2019fdr, AT2019aalc is also accompanied by a dust echo. The full sample was found to be spatially and temporally correlated with IceCube high-energy neutrino alerts at the 3.6$\sigma$ level \citep{van_Velzen_Stein_Gilfanov_Kowalski_Hayasaki_Reusch_Yao_Garrappa_Franckowiak_Gezari_et_al._2024}.
The arrival times of the associated neutrinos exhibit significant delays with respect to the peak of the OUV emission. For AT2019dsg, the neutrino was detected approximately 150 days after the optical peak, while delays of $\sim300$ days and $\sim150 $ days were reported for AT2019fdr and AT2019aalc, respectively. These delayed neutrino emissions have been proposed to arise from interactions with IR photons whose dust echoes peak at $\sim239$, $\sim155$, and $\sim78$ days after the OUV peak for AT2019dsg, AT2019fdr, and AT2019aalc, respectively \citep{Winter_Lunardini_2023}. An additional search using IceCube data and a subset of accretion flares exhibiting strong dust echoes did not reveal any significant excess over the background expectation \citep{necker2023searchhighenergyneutrinostdelike}. In addition, improved directional reconstruction with IceCube track alert catalogs has weakened the previously reported associations, rendering the candidates no longer consistent with the neutrino localization regions \citep{Zegarelli_Franckowiak_Sommani_Valtonen-Mattila_Yuan_2025}.

Because dust reprocessing is largely agnostic to the triggering mechanism, dust echoes provide access to a broader population of transient accretion events, including obscured systems where the primary optical or UV emission may be attenuated. If high-energy neutrino production is linked to interactions with IR photons from dust echoes \citep{Winter_Lunardini_2023}, IR-selected samples may provide a direct probe of the neutrino-emitting population. To fully exploit the available neutrino data, it is therefore essential to consider source samples that span the full IceCube data-taking period, that do not rely on optical detections, and that provide broad sky coverage. In this work, we perform a stacking analysis of TDE candidates selected from the largest sample of IR accretion flares compiled to date by \citet{Necker_2025}, to search for a cumulative high-energy neutrino signal from this source population.

The paper is organized as follows. In Sect.~\ref{sec:data} we describe the IR source catalog and IceCube dataset. The analysis method is presented in Sect.~\ref{sec:method}, followed by the results in Sect.~\ref{sec:results}. We discuss the implications of our findings in Sect.~\ref{sec:discussion} and summarize our conclusions in Sect.~\ref{sec:conclusion}.

%__________________________________________________________________

\section{Source catalog and data selection}
\label{sec:data}

Starting from a sample of $\mathcal{O}(10^6)$ variable IR light curves, \cite{Necker_2025} identified dust echoes associated with transient accretion activity using archival WISE data \citep{Wright_Eisenhardt_Mainzer_Ressler_Cutri_Jarrett_Kirkpatrick_Padgett_McMillan_Skrutskie_et_al._2010, Mainzer_Bauer_Cutri_Grav_Masiero_Beck_Clarkson_Conrow_Dailey_Eisenhardt_et_al._2014} and compiled the Flaires catalog, consisting of 823 dust-echo-like flares. The IR selection preferentially identifies luminous and dust-rich transients, while the sky coverage is limited by the parent optical and IR surveys used in the catalog construction. The Flaires catalog employs variability-based selection criteria to isolate transient nuclear flares and uses cross-matching procedures to remove likely synchrotron-dominated sources. For the present analysis, we exclude objects with AGN classifications in the catalog to focus on the TDE-candidate population. From the parent accretion-flare catalog, we retain only sources with estimated dust properties, yielding a final sample of 528 infrared flares.

This analysis uses the IceCube Northern Tracks sample, comprising ten years (2011--2021) of through-going muon track data \citep{nuflux_2022, nuflux_2025}, collected with the 86-string (IC86) detector configuration \citep{Aartsen_Ackermann_Adams_Aguilar_Ahlers_Ahrens_Altmann_Andeen_Anderson_Ansseau_et_al._2017}, completed in December 2010. This sample consists of elongated Cherenkov light signatures known as tracks left by muons produced in charged-current $\nu_\mu$ interactions. Track events provide sub-degree angular resolution and are therefore preferred for point-source searches.  In contrast, charged-current $\nu_e$ and $\nu_\tau$ interactions, as well as neutral current interactions of all three neutrino flavors, produce approximately spherically symmetric light signatures known as cascades, which have larger directional uncertainties than track events \citep{Aartsen_cascade_al._2017}. The Northern Tracks sample comprises upgoing events, i.e. events arriving always from below the horizon at the South Pole. For this event selection, the Earth acts as an effective shield against the atmospheric-muon background, removing most of the downgoing atmospheric muons, but limiting the sensitivity to declinations $\delta > -5^\circ$. We therefore consider only sources in this declination range (which we simply refer to as the "Northern Sky" in the following) occurring during the data-taking period covered by the full IceCube sample.

\begin{figure}[h!]
\centering
    \centering
    \includegraphics[width=\linewidth]{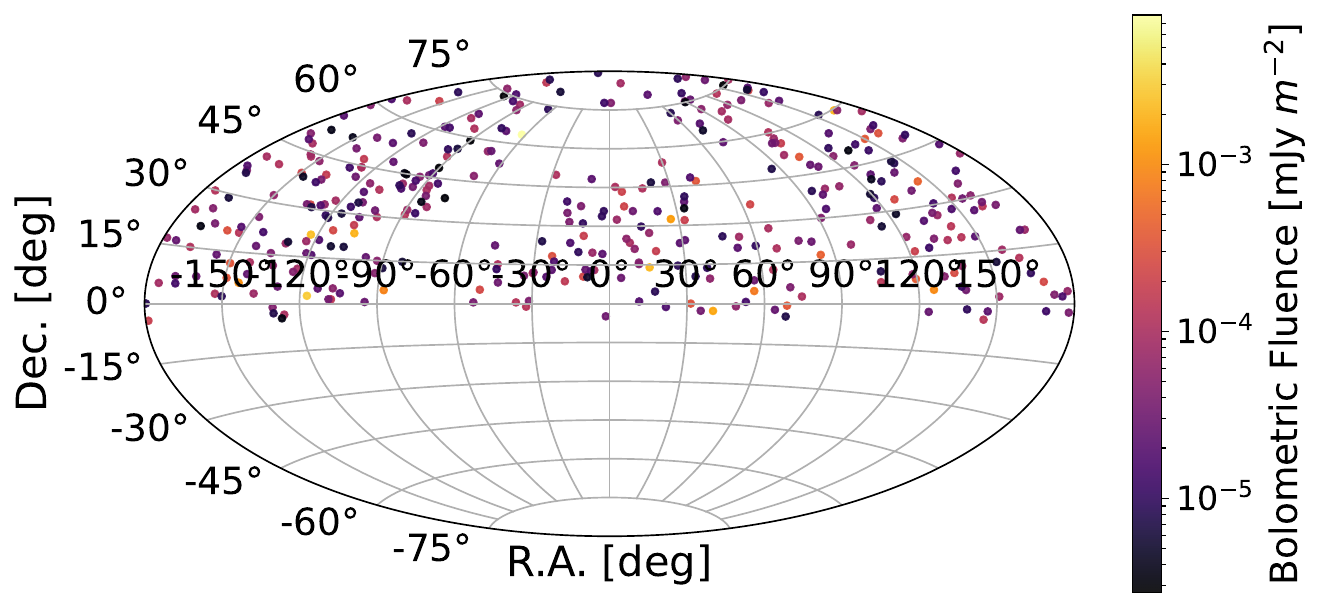}
    \label{fig:northern}
\caption{Sky map in equatorial coordinates of the 393 sources in the Northern Sky sample used in this analysis. The color map shows the total bolometric fluence of each source, used as a weight in the likelihood method (see Sect.~\ref{sec:method}). The reduced source density along the Galactic plane reflects the limited sky coverage of the parent optical and IR surveys used to construct the Flaires catalog, whose classification of sources as stars or galaxies becomes unreliable in crowded stellar fields \citep{Necker_2025}.}
\label{fig:skymap}
\end{figure}

None of the previously reported TDE candidates associated with IceCube neutrino alerts, based on the previously available localization contours, are included. The source AT2019fdr is excluded because its comparatively faint, higher-redshift host galaxy is not in the parent galaxy sample used to build the catalog, and AT2019aalc failed the variability-selection criteria designed to distinguish isolated dust echoes from stochastic AGN variability \citep{Necker_2025}. Although AT2019dsg is recovered by the Flaires pipeline and classified as a “dust-echo-like” flare, it is excluded from the stacking analysis to avoid bias from an a posteriori source selection since it has previously been discussed in connection with a high-energy neutrino event. %and is used as a benchmark example in the IR neutrino production model of \cite{Winter_Lunardini_2023}. 
After applying all selection criteria, the final sample consists of 393 sources in the Northern Sky, as shown in Fig.~\ref{fig:skymap}.

%
%______________________________________________________________

\section{Analysis method}
\label{sec:method}

To search for correlations between IR flares and high-energy neutrinos, we perform a time-dependent stacking analysis of the Flaires catalog using the unbinned maximum-likelihood method of \citet{Braun_Dumm_De_Palma_Finley_Karle_Montaruli_2008, Braun_Baker_Dumm_Finley_Karle_Montaruli_2010} applied to 10 years of Northern-sky muon-track data. To determine whether each event originates from the selected IR flares or from the background, this approach employs probability density functions (PDFs). The PDFs for both signal and background are factorized into three parts: spatial, energy, and temporal. The background PDF is built from Monte Carlo (MC) simulations of atmospheric background. The atmospheric neutrino component is modeled using the software MCEq \citep{Fedynitch_Engel_Gaisser_Riehn_Stanev_2015}, assuming the GST cosmic-ray flux model \citep{Gaisser_Stanev_Tilav_2013} and the SIBYLL 2.3c hadronic interaction model \citep{Fedynitch_Riehn_Engel_Gaisser_Stanev_2019}. The atmospheric muon component is separately estimated using CORSIKA air-shower simulation \citep{Heck_Knapp_Capdevielle_Schatz_Thouw_1998}, assuming the GaisserH4a cosmic-ray flux model \citep{Gaisser_2012} and the same SIBYLL 2.3c hadronic interaction model. For the signal, the spatial PDF describes the probability density of reconstructing an event at a given angular separation from the source position. The energy component of the signal PDF is generated using MC-simulated astrophysical events. For these simulations, a power-law injection model $E_{\nu}^{-\gamma}$ is assumed, and each MC event is assigned a weight reflecting this astrophysical spectrum. The temporal signal PDF is modeled as a one-year box window ending at the IR peak time of each source. Since the IR emission is expected to be delayed relative to the primary accretion flare due to dust reprocessing, the IR peak provides a reference time available for all sources in the Flaires sample. This choice is intended to capture neutrino emission occurring prior to the maximum of the dust echo. The signal PDF for source $j$ is therefore given by:

\begin{equation}
    \mathcal{S}_j(\nu_i, \gamma) = \mathcal{S}_j^{\rm spatial}(\nu_i)\times\mathcal{S}^{\rm energy}(\nu_i, \gamma)\times\mathcal{S}_j^{\rm time}(t_i)\,,
\end{equation}
where $\nu_i \equiv (\alpha_i, \delta_i, E_i, t_i)$ denotes the reconstructed direction, angular uncertainty, energy, and arrival time of the $i_{\rm th}$ neutrino. A common power-law spectrum with spectral index $\gamma$ is assumed for all catalog sources. In the analysis, the likelihood is evaluated using the weighted sum of the individual source contributions $\mathcal{S} (\nu_i, \gamma) = \sum_{j=1}^M w_j \mathcal{S}_j(\nu_i, \gamma)$, where $M$ is the number of sources, $\mathcal{S}_j(\nu_i, \gamma)$ is the signal PDF associated with source $j$, and $w_j$ represents the relative weight assigned to each source. In this work, the source weights are determined using the bolometric fluence of each IR flare\footnote{An accepted erratum to the Flaires catalog corrects the luminosity and fluence estimates by a factor of 2 \citep{Necker_Graikou_Kowalski_Franckowiak_Nordin_Pernice_Van_Velzen_Veres_2026}. Since this analysis uses fluence only as a relative source weight, the correction is not expected to affect the results.}, under the assumption that the expected neutrino emission scales proportionally with the total radiated energy. The bolometric fluence is derived from a blackbody fit to the IR pho\-to\-me\-try at each epoch of the source's light curve, integrated over the flare duration and converted to a fluence using the source luminosity distance \citep{Necker_2025}.
%This quantity was estimated by integrating the bolometric luminosity over the flare duration, using discrete WISE observations. 
The likelihood function then becomes:
\begin{equation}
L(n_s, \gamma) = \prod_{i=1}^{N} \left(\frac{n_s}{N} \sum_{j=1}^{M}w_j\, \mathcal{S}_j(\nu_{i}, \gamma) + \left(1 - \frac{n_{s}}{N}\right) \mathcal{B}(\nu_i) \right)\,.
\end{equation}
The first product runs over all the ${\displaystyle N}$ neutrino events in the dataset, where ${\mathcal{S}_j}(\nu_i,\gamma)$ and ${\mathcal{B}}(\nu_i)$ represent, respectively, the signal and background PDFs for the $i_{\mathrm{th}}$ neutrino and $j_{\mathrm{th}}$ source. The parameter $n_s$ represents the estimated number of signal events in the dataset, such that $n_s/N$ and $1-n_s/N$ correspond to the signal and background fractions, respectively.

To quantify the presence of a potential signal excess, we define the test statistic (TS) as the logarithmic likelihood ratio between the background and signal hypotheses:
\begin{equation}\label{eq:TS}
    \mathrm{TS} = -2 \log \frac{L(n_s=0)}{L(\hat{n_s}, \hat{\gamma})}\,,
\end{equation}
where $L(n_s=0)$ corresponds to the likelihood of observing background and \(L(\hat{n_s},\hat{\gamma})\) is the likelihood maximized over the number of signal events $\hat{n}_s$ and the spectral index $\hat{\gamma}$. The significance of the observed TS is evaluated by comparing the TS obtained from data to the TS distribution derived from pseudo-experiments generated under the background-only hypothesis. We then compute the corresponding p-value, which quantifies how compatible the observed TS is with the background-only scenario.
%
%______________________________________________________________

\section{Results}
\label{sec:results}
We find no significant correlation between 393 IR flares in the Northern Sky and 10 years of IceCube through-going track data. The observed TS is 0.0, coinciding with the median of the background distribution, indicating no significant excess over the background expectation.
The likelihood fit yields a best-fit number of signal events of  \(\hat{n_s} = 0.0\), corresponding to a local p-value of 0.998.

To assess the performance of the analysis, we first evaluate its sensitivity to neutrino emission from the source catalog. For a stacking search, the sensitivity is defined as the common flux normalization of the stacked source population for which the TS exceeds the median of the background-only distribution in 90\% of signal-injected pseudo-experiments \citep{Braun_Dumm_De_Palma_Finley_Karle_Montaruli_2008, Aartsen_Ackermann_Adams_Aguilar_Ahlers_Ahrens_Alispach_Andeen_Anderson_Ansseau_et_al._2020}. The pseudo-experiments are generated by injecting simulated neutrino signals from all catalog sources into background-only trials. The relative contribution of each source is determined by its source weight and the declination-dependent detector effective area, which accounts for the declination dependence of the detector response. For each assumed spectral index, the injected signal strength is varied until the sensitivity criterion is satisfied. We repeat this procedure for three spectral indices, $\gamma=1.0$, $2.0$, and $2.5$, to probe different neutrino-emission scenarios, with $\gamma=2.0$ and $2.5$ representing commonly adopted benchmark spectra for astrophysical neutrino emission. In particular, we consider a hard spectral index of $\gamma=1$, motivated by theoretical models in which neutrinos are produced through interactions between ultra-high-energy (UHE) protons and IR photons from dust echoes \citep{Winter_Lunardini_2023}. Such scenarios predict hard neutrino spectra with emission extending predominantly to the highest energies. In addition, the previous stacking analysis by \citet{necker2023searchhighenergyneutrinostdelike} of accretion flares exhibiting strong dust echoes derived ULs on the source flux for different spectral indices. The results showed that a spectral index harder than $\gamma=2$ is required to explain the previously reported TDE-neutrino alert associations without predicting an excess in the full stacked sample, further motivating the choice of testing $\gamma=1$. 

In the absence of a significant signal excess, the corresponding 90\% confidence level C.L. upper limits (ULs) on the stacked flux normalization are derived following the same procedure used for the sensitivity. Since the observed TS is $0$, the ULs coincide with the sensitivities. The Flaires catalog 90\% C.L. ULs on the single-flavor ($\nu_\mu+\bar{\nu}_\mu$) flux from the Northern Sky are summarized in Table~\ref{tab:UL} for the three assumed values of the spectral index. For comparison across the different spectral assumptions, the ULs are expressed as the flux normalization at a common reference energy of 100 TeV.

\begin{figure}[h!]
\centering
    \centering
    \includegraphics[width=\linewidth]{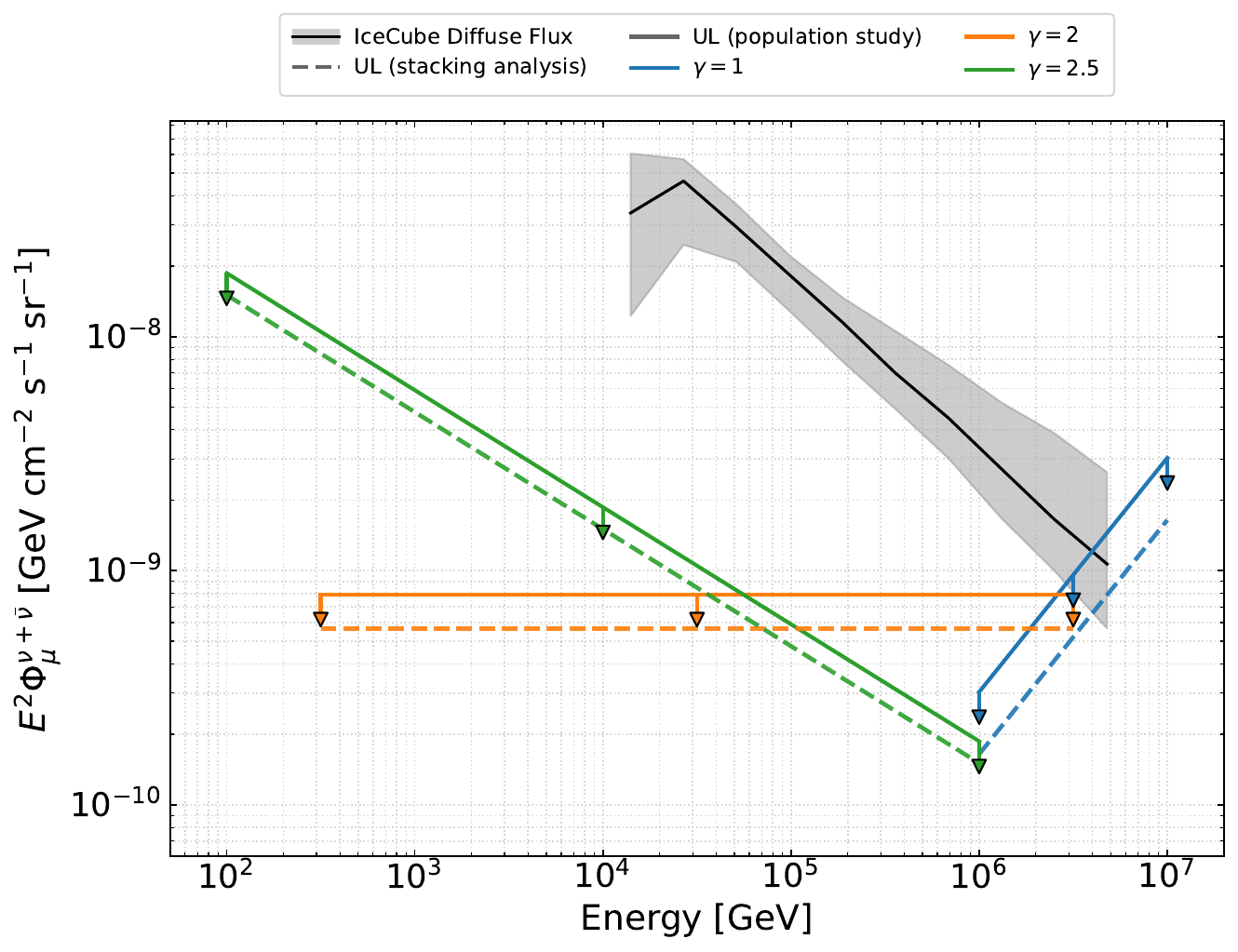}
\caption{Upper limits on the sky-averaged muon-neutrino flux from infrared flares in the Northern Sky are shown for three assumed spectral indices ($\gamma = 1, 2$ and $2.5$). Dashed lines correspond to the ULs obtained from the stacking analysis, while solid lines show the ULs derived from the population study. For each spectral index, the corresponding UL is shown only over the energy range in which the analysis is most sensitive. The gray band indicates the 95$\%$ confidence interval of the diffuse astrophysical neutrino flux measured by IceCube \citep{Abbasi_Ackermann_Adams_Agarwalla_Aguilar_Ahlers_Alameddine_Ali_Amin_Andeen_et_al._2026_break}. The diffuse flux is shown under the assumption of single-flavor normalization.}
\label{fig:pop_study}
\end{figure}

To constrain the contribution of IR-selected TDEs to the diffuse neutrino flux, we adopt a simplified standard-candle population model, in which we assume all sources share the same intrinsic neutrino spectrum and luminosity. Since no luminosity function is currently available for the Flaires TDE population due to uncertainties in the completeness of the parent sample \citep{Necker_2025}, the resulting population constraints should be regarded as model-dependent. To interpret the stacking ULs in terms of a source population, we convert the catalog UL on the common flux normalization into an effective UL for a representative source, UGC 11487, corresponding to the host galaxy of the brightest flare in our sample.  This conversion, performed independently for each of the three assumed spectral indices, is obtained by scaling the stacking UL with the normalized relative weight of the representative source, yielding the corresponding flux upper limit for that source. We additionally account for the source declination and flare duration to incorporate the IceCube effective area. The resulting effective flux UL is then converted into the corresponding neutrino luminosity of the representative source, using its luminosity distance and the redshift correction appropriate for the assumed power-law spectrum. This luminosity is adopted as the standard-candle luminosity of the source population, and combined with the assumed volumetric rate density and its redshift evolution to derive the diffuse-flux UL. As the Flaires catalog does not provide a volumetric rate estimate, we adopt the local volumetric TDE rate estimated by \cite{masterson2024new}, $(1.3 \pm 0.2)\times10^{-7} \mathrm{Mpc}^{-3} \mathrm{yr}^{-1}$. This estimate is based on a high-purity ``gold'' sample of IR-selected TDEs and corresponds to a per-galaxy rate of $(2.0 \pm 0.3)\times10^{-5} \mathrm{galaxy}^{-1} \mathrm{yr}^{-1}$, consistent with recent measurements from optical TDE surveys \citep{Yao_Ravi_Gezari_Van_Velzen_Lu_Schulze_Somalwar_Kulkarni_Hammerstein_Nicholl_et_al._2023}. Moreover, this galaxy rate is comparable to that inferred for the Flaires population. In addition, a substantial fraction of the \cite{masterson2024new} sample is also contained in the Flaires catalog, including our reference source. Given the overlap between the two populations and the absence of a dedicated volumetric rate measurement for Flaires TDEs, we adopt this value as the local rate normalization in our population study. For the redshift evolution, we assume a model derived from numerical data from \cite{sun2015extragalactic}. The resulting diffuse-flux ULs obtained under these assumptions are shown in Fig.~\ref{fig:pop_study}. 

The sky-averaged ULs are compared to the diffuse astrophysical neutrino flux measured by IceCube \citep{Abbasi_Ackermann_Adams_Agarwalla_Aguilar_Ahlers_Alameddine_Ali_Amin_Andeen_et_al._2026_break} to estimate the maximum fractional contribution of IR-selected TDEs to this flux. For each assumed spectral index, this contribution is obtained by comparing our population-level flux UL to the measured diffuse flux, both evaluated at the same reference energy of 100 TeV. Assuming that the Flaires catalog is predominantly composed of TDEs, we constrain their contribution to the observed diffuse muon-neutrino flux to be below $0.14\% $, $3.5 \%$, $2.7 \%$ at 100 TeV for $\gamma$ = 1, 2, 2.5, respectively. These estimates are summarized in Table~\ref{tab:UL} and should be interpreted as ULs on the diffuse neutrino contribution from the source population, given the assumptions on the source evolution and the neutrino emission model.

\begin{table}[h!]
    \renewcommand{\arraystretch}{1.3}
    \centering
    \begin{tabular}{|c c c c|}
        \hline
        & $\gamma = 1$ & $\gamma = 2$ & $\gamma = 2.5$ \\
        \hline
        90\% C.L. U.L.
        & $1.65 \times 10^{-11}$ 
        & $5.67 \times 10^{-10}$ 
        & $4.76 \times 10^{-10}$ \\
        Fraction U.L.
        & $0.14\%$
        & $3.5\%$
        & $2.7\%$ \\
        \hline
    \end{tabular}
    \caption{Upper limits (ULs) at 90\% C.L. on the single-flavor ($\nu_\mu+\bar{\nu}_\mu$) flux, evaluated at 100 TeV, and corresponding upper limits on the contribution of the TDE population to the observed diffuse muon-neutrino flux. The flux limits are expressed as $E^{2}\Phi(E)$ in units of $\mathrm{GeV\,cm^{-2}\,s^{-1}\,sr^{-1}}$.}
    \label{tab:UL}
\end{table}

\section{Discussions}
\label{sec:discussion}

This stacking analysis reveals no statistically significant correlation between IR flares and high-energy neutrinos and indicates that IR-selected TDEs do not account for a dominant fraction of the observed IceCube diffuse neutrino flux under the adopted assumptions. Our results are consistent with previous searches for neutrino emission from TDEs and accretion-powered transients. In particular, \cite{masterson2024new} found no significant spatial or temporal correlations between their "gold" sample of IR-selected TDEs and IceCube astrophysical neutrino candidates. Similarly, no significant excess has been found in stacking searches correlating IceCube data with a broader population of nuclear accretion flares \citep{necker2023searchhighenergyneutrinostdelike}. Stacking analyses of optically selected TDE samples have likewise yielded null results, constraining their contribution to the diffuse astrophysical neutrino flux to at most a few tens of percent \citep{Stein_IceCube_Collaboration_2019, Mukhopadhyay_Wusinich_Murase_2026}.
Furthermore, a recent study testing TDEs identified in the X-ray and OUV bands and IceCube data yielded null results and also reports ULs on the contribution of jetted and non-jetted TDEs to the diffuse neutrino flux observed by IceCube \citep{Abbasi_Ackermann_Adams_Aguilar_Ahlers_Alameddine_Ali_Amin_Andeen_Arguelles_et_al._2026}.

Even though theoretical models have proposed that TDEs with prominent dust echoes could contribute substantially to the diffuse neutrino flux \citep{Winter_Lunardini_2023}, our results place constraints on the cumulative neutrino emission from this population. However, a detailed comparison requires adopting the specific spectral and population assumptions of the model, and our limits should be interpreted in light of the selection effects of the source catalog as well. The Flaires selection pipeline achieves an estimated completeness of roughly 30$\%$ to $50\%$ when compared to other IR flare samples, which is the result of more stringent filters to ensure sample purity against stochastic AGN activity \citep{Necker_2025}. In addition, the analysis is restricted to Northern-sky sources and therefore does not fully sample the underlying IR-flare population represented by the Flaires catalog. An additional source of model dependence arises from the assumption of a power-law neutrino spectrum in the likelihood. While this choice enables a model-independent interpretation, realistic TDE scenarios may produce more complex spectral shapes and temporal evolution, thereby providing more stringent and physically informative constraints. The population study results introduce additional uncertainties associated with the adopted local volumetric rate and the redshift evolution of the source sample. While \citet{Necker_2025} find evidence for a redshift evolution of the flare rate that declines more steeply toward high redshift than expected from models based purely on the density of supermassive black holes \citep{sun2015extragalactic, Kochanek_2016}, they do not provide an analytical parameterization. A redshift evolution that decreases more rapidly with increasing redshift would reduce the contribution from distant sources and therefore lead to more stringent constraints on the diffuse neutrino flux. Consequently, the inferred population contribution should be regarded as model-dependent. 
To assess the impact of the population evolution, we additionally test a simplified cosmological parameterization of the source volumetric rate as a function of cosmological redshift, $f(z)=(1+z)^{-m}$, where $z$ is the redshift and $m$ controls how rapidly the source rate decreases with increasing redshift. By adopting the same methodology reported in Sec. ~\ref{sec:results}, we show the corresponding ULs in Appendix~\ref{appendix} (Fig.~\ref{fig:model_pop_study}).

In their work, \citet{Winter_Lunardini_2023} compute the diffuse neutrino flux expected from the full cosmological population of TDEs and compare it to IceCube data. This prediction is obtained by integrating the neutrino emission, calibrated on the three individually reported neutrino-TDE associations (AT2019dsg, AT2019fdr, and AT2019aalc), over the black hole mass function and cosmic history. The resulting flux depends on the product of the dissipation efficiency $\varepsilon_{diss}$, the fraction of the accretion power converted into accelerated protons, and $\eta$, the fraction of the TDE population in which the mechanism is active. We extrapolate their flux normalization $\phi_{model}$ corresponding to $\eta\varepsilon_{diss}=1$, and compare this to our own population-level flux normalization $\phi_0$ for $\gamma=1$, rescaled to the same reference energy of 100 TeV. Assuming that the mechanism is active in every TDE ($\eta=1$), our stacking result can be translated into an upper limit on the dissipation efficiency:
\begin{equation}
    \varepsilon_{diss} < \frac{\phi_0(100\,\mathrm{TeV})}{\phi_{model}} \approx 2.8\times10^{-3}\,,
\end{equation}
approximately a factor of two more stringent than the corresponding limit derived by \citet{Necker:2025pii} from a 29-source stacking analysis ($\varepsilon_{diss}<6\times10^{-3}$) of accretion flares. This result is one to two orders of magnitude below the benchmark values assumed by \citet{Winter_Lunardini_2023} to reproduce the observed diffuse flux. This places significant constraints on the scenario in which the IR mechanism operates universally at the efficiency proposed. In addition, since our sample does not include the three sources used to calibrate the IR model, this comparison constitutes an independent test of whether the calibrated mechanism extends to a broader TDE population.

Our result is also consistent with the revised IceCube alert reconstruction. The TDE candidates that originally motivated this analysis were identified through spatial and temporal coincidence with IceCube alerts. However, updated event reconstructions have led to a revised alert catalog (IceCat-2) \citep{Abbasi_Ackermann_Adams_Agarwalla_Aguilar_Ahlers_Alameddine_Ali_Amin_Andeen_et_al._2025}, in which these sources are no longer consistent with the neutrino localization regions.

\section{Conclusion}
\label{sec:conclusion}

In this work, we performed a stacking search for neutrino emission from a sample of IR-selected TDE candidates using 10 years of IceCube through-going tracks. The analysis was motivated by recent correlations between dust-obscured accretion flares and high-energy neutrino alerts \citep{Stein_Velzen_Kowalski_Franckowiak_Gezari_Miller-Jones_Frederick_Sfaradi_Bietenholz_Horesh_et_al, reusch2022candidate, van_Velzen_Stein_Gilfanov_Kowalski_Hayasaki_Reusch_Yao_Garrappa_Franckowiak_Gezari_et_al._2024}, and by theoretical models predicting the observed delayed arrival time of neutrinos relative to the OUV peak \citep{Winter_Lunardini_2023}. No statistically significant neutrino excess was found.

In the absence of a significant signal, we derive ULs on the contribution of this source class to the diffuse astrophysical neutrino flux, based on assumptions on the source evolution and population properties. The results remain consistent with previous null results from stacking and correlation studies of TDEs, providing stronger constraints on the contribution of these sources to the neutrino diffuse flux observed by IceCube for different spectral assumptions on the emitting sources.

Future improvements may come from larger and more complete samples of dust-obscured accretion flares and TDE candidates, extended neutrino datasets and improved statistical inference frameworks. Time-domain surveys such as the Vera C. Rubin Observatory Legacy Survey of Space and Time (LSST) \citep{Ivezic_Kahn_Tyson_Abel_Acosta_Allsman_Alonso_AlSayyad_Anderson_Andrew_et_al._2019} are expected to drastically increase the number of detected TDE candidates and improve constraints on the redshift evolution of the TDE population \citep{Karmen_Gezari_Norman_Guolo_2026}. Recent developments in the IceCube detector will further improve the sensitivity of future analyses. The recent implementation of the IceCube Upgrade has fully commissioned 5 additional, more densely instrumented strings at the center of the array, providing enhanced sensitivity to GeV-scale neutrinos as well as improved calibration of the optical properties of the surrounding ice \citep{IceCube_Collaboration_2025}. Looking further ahead, the next-generation detector, IceCube-Gen2, will provide enhanced sensitivity to high-energy neutrino fluxes \citep{Aartsen_Abbasi_Ackermann_Adams_Aguilar_Ahlers_Ahrens_Alispach_Allison_Amin_et_al._2021}. Future analyses could also benefit from Bayesian population-inference techniques, which may allow uncertainties in the source evolution, luminosity distribution, and neutrino-emission properties to be incorporated directly into the stacking framework \citep{bartos2025bayesian}.

\begin{acknowledgements}
The IceCube collaboration acknowledges the significant contributions to this manuscript from Teresa Pernice.
The authors gratefully acknowledge the support from the following agencies and institutions:
USA {\textendash} U.S. National Science Foundation-Office of Polar Programs,
U.S. National Science Foundation-Physics Division,
U.S. National Science Foundation-EPSCoR,
U.S. National Science Foundation-Office of Advanced Cyberinfrastructure,
Wisconsin Alumni Research Foundation,
Center for High Throughput Computing (CHTC) at the University of Wisconsin{\textendash}Madison,
Open Science Grid (OSG),
Partnership to Advance Throughput Computing (PATh),
Advanced Cyberinfrastructure Coordination Ecosystem: Services {\&} Support (ACCESS),
Frontera and Ranch computing project at the Texas Advanced Computing Center,
U.S. Department of Energy-National Energy Research Scientific Computing Center,
Particle astrophysics research computing center at the University of Maryland,
Michigan State University,
Astroparticle physics computational facility at Marquette University,
NVIDIA Corporation,
and Google Cloud Platform;
Belgium {\textendash} Funds for Scientific Research (FRS-FNRS and FWO),
FWO Odysseus and Big Science programmes,
and Belgian Federal Science Policy Office (Belspo);
Germany {\textendash} Bundesministerium f{\"u}r Forschung, Technologie und Raumfahrt (BMFTR),
Deutsche Forschungsgemeinschaft (DFG),
Helmholtz Alliance for Astroparticle Physics (HAP),
Initiative and Networking Fund of the Helmholtz Association,
Deutsches Elektronen Synchrotron (DESY),
and High Performance Computing cluster of the RWTH Aachen;
Sweden {\textendash} Swedish Research Council,
Swedish Polar Research Secretariat,
National Academic Infrastructure for Supercomputing in Sweden (NAISS),
and Knut and Alice Wallenberg Foundation;
European Union {\textendash} EGI Advanced Computing for research;
Australia {\textendash} Australian Research Council;
Canada {\textendash} Natural Sciences and Engineering Research Council of Canada,
Calcul Qu{\'e}bec, Compute Ontario, Canada Foundation for Innovation, WestGrid, and Digital Research Alliance of Canada;
Denmark {\textendash} Villum Fonden, Carlsberg Foundation, and European Commission;
New Zealand {\textendash} Marsden Fund;
Japan {\textendash} Japan Society for Promotion of Science (JSPS), Ministry of Education, Culture, Sports, Science and Technology (MEXT), and Institute for Global Prominent Research (IGPR) of Chiba University;
Korea {\textendash} National Research Foundation of Korea (NRF);
Switzerland {\textendash} Swiss National Science Foundation (SNSF).
\end{acknowledgements}

%-------------------------------------------------------------------
\bibliographystyle{bibtex/aa}
\bibliography{reference}

@article{Winter_Lunardini_2023, title={Interpretation of the observed neutrino emission from three Tidal Disruption Events}, volume={948}, ISSN={0004-637X, 1538-4357}, DOI={10.3847/1538-4357/acbe9e}, note={arXiv:2205.11538 [astro-ph]}, number={1}, journal={The Astrophysical Journal}, author={Winter, Walter and Lunardini, Cecilia}, year={2023}, month=may, pages={42} }

@article{Stein_Velzen_Kowalski_Franckowiak_Gezari_Miller-Jones_Frederick_Sfaradi_Bietenholz_Horesh_et_al, title={A tidal disruption event coincident with a high-energy neutrino}, volume={5}, ISSN={2397-3366}, DOI={10.1038/s41550-020-01295-8}, number={5}, journal={Nature Astronomy}, author={Stein, Robert and Velzen, Sjoert van and Kowalski, Marek and Franckowiak, Anna and Gezari, Suvi and Miller-Jones, James C. A. and Frederick, Sara and Sfaradi, Itai and Bietenholz, Michael F. and Horesh, Assaf and Fender, Rob and Garrappa, Simone and Ahumada, Tomás and Andreoni, Igor and Belicki, Justin and Bellm, Eric C. and Böttcher, Markus and Brinnel, Valery and Burruss, Rick and Cenko, S. Bradley and Coughlin, Michael W. and Cunningham, Virginia and Drake, Andrew and Farrar, Glennys R. and Feeney, Michael and Foley, Ryan J. and Gal-Yam, Avishay and Golkhou, V. Zach and Goobar, Ariel and Graham, Matthew J. and Hammerstein, Erica and Helou, George and Hung, Tiara and Kasliwal, Mansi M. and Kilpatrick, Charles D. and Kong, Albert K. H. and Kupfer, Thomas and Laher, Russ R. and Mahabal, Ashish A. and Masci, Frank J. and Necker, Jannis and Nordin, Jakob and Perley, Daniel A. and Rigault, Mickael and Reusch, Simeon and Rodriguez, Hector and Rojas-Bravo, César and Rusholme, Ben and Shupe, David L. and Singer, Leo P. and Sollerman, Jesper and Soumagnac, Maayane T. and Stern, Daniel and Taggart, Kirsty and van Santen, Jakob and Ward, Charlotte and Woudt, Patrick and Yao, Yuhan}, year={2021}, month=may, pages={510–518} }

@article{Zegarelli_Franckowiak_Sommani_Valtonen-Mattila_Yuan_2025, title={IceCat-2: Updated IceCube Event Catalog of Alert Tracks}, url={http://arxiv.org/abs/2507.06176}, DOI={10.48550/arXiv.2507.06176}, note={arXiv:2507.06176 [astro-ph]}, number={arXiv:2507.06176}, publisher={arXiv}, author={Zegarelli, Angela and Franckowiak, Anna and Sommani, Giacomo and Valtonen-Mattila, Nora and Yuan, Tianlu}, year={2025}, month=jul }

@article{Aartsen_Ackermann_Adams_Aguilar_Ahlers_Ahrens_Altmann_Andeen_Anderson_Ansseau_et_al._2017, title={The IceCube Neutrino Observatory: instrumentation and online systems}, volume={12}, DOI={10.1088/1748-0221/12/03/P03012}, number={03}, journal={Journal of Instrumentation}, author={Aartsen, M.G. and Ackermann, M. and Adams, J. and Aguilar, J.A. and Ahlers, M. and Ahrens, M. and Altmann, D. and Andeen, K. and Anderson, T. and Ansseau, I. and Anton, G. and Archinger, M. and Argüelles, C. and Auer, R. and Auffenberg, J. and Axani, S. and Baccus, J. and Bai, X. and Barnet, S. and Barwick, S.W. and Baum, V. and Bay, R. and Beattie, K. and Beatty, J.J. and Tjus, J. Becker and Becker, K.-H. and Bendfelt, T. and BenZvi, S. and Berley, D. and Bernardini, E. and Bernhard, A. and Besson, D.Z. and Binder, G. and Bindig, D. and Bissok, M. and Blaufuss, E. and Blot, S. and Boersma, D. and Bohm, C. and Börner, M. and Bos, F. and Bose, D. and Böser, S. and Botner, O. and Bouchta, A. and Braun, J. and Brayeur, L. and Bretz, H.-P. and Bron, S. and Burgman, A. and Burreson, C. and Carver, T. and Casier, M. and Cheung, E. and Chirkin, D. and Christov, A. and Clark, K. and Classen, L. and Coenders, S. and Collin, G.H. and Conrad, J.M. and Cowen, D.F. and Cross, R. and Day, C. and Day, M. and de André, J.P.A.M. and Clercq, C. De and Rosendo, E. del Pino and Dembinski, H. and Ridder, S. De and Descamps, F. and Desiati, P. and de Vries, K.D. and de Wasseige, G. and de With, M. and DeYoung, T. and D{\'i}az-V{\'e}lez, J.C. and di Lorenzo, V. and Dujmovic, H. and Dumm, J.P. and Dunkman, M. and Eberhardt, B. and Edwards, W.R. and Ehrhardt, T. and Eichmann, B. and Eller, P. and Euler, S. and Evenson, P.A. and Fahey, S. and Fazely, A.R. and Feintzeig, J. and Felde, J. and Filimonov, K. and Finley, C. and Flis, S. and Fösig, C.-C. and Franckowiak, A. and Frère, M. and Friedman, E. and Fuchs, T. and Gaisser, T.K. and Gallagher, J. and Gerhardt, L. and Ghorbani, K. and Giang, W. and Gladstone, L. and Glauch, T. and Glowacki, D. and Glüsenkamp, T. and Goldschmidt, A. and Gonzalez, J.G. and Grant, D. and Griffith, Z. and Gustafsson, L. and Haack, C. and Hallgren, A. and Halzen, F. and Hansen, E. and Hansmann, T. and Hanson, K. and Haugen, J. and Hebecker, D. and Heereman, D. and Helbing, K. and Hellauer, R. and Heller, R. and Hickford, S. and Hignight, J. and Hill, G.C. and Hoffman, K.D. and Hoffmann, R. and Hoshina, K. and Huang, F. and Huber, M. and Hulth, P.O. and Hultqvist, K. and In, S. and Inaba, M. and Ishihara, A. and Jacobi, E. and Jacobsen, J. and Japaridze, G.S. and Jeong, M. and Jero, K. and Jones, A. and Jones, B.J.P. and Joseph, J. and Kang, W. and Kappes, A. and Karg, T. and Karle, A. and Katz, U. and Kauer, M. and Keivani, A. and Kelley, J.L. and Kemp, J. and Kheirandish, A. and Kim, J. and Kim, M. and Kintscher, T. and Kiryluk, J. and Kitamura, N. and Kittler, T. and Klein, S.R. and Kleinfelder, S. and Kleist, M. and Kohnen, G. and Koirala, R. and Kolanoski, H. and Konietz, R. and Köpke, L. and Kopper, C. and Kopper, S. and Koskinen, D.J. and Kowalski, M. and Krasberg, M. and Krings, K. and Kroll, M. and Krückl, G. and Krüger, C. and Kunnen, J. and Kunwar, S. and Kurahashi, N. and Kuwabara, T. and Labare, M. and Laihem, K. and Landsman, H. and Lanfranchi, J.L. and Larson, M.J. and Lauber, F. and Laundrie, A. and Lennarz, D. and Leich, H. and Lesiak-Bzdak, M. and Leuermann, M. and Lu, L. and Ludwig, J. and Lünemann, J. and Mackenzie, C. and Madsen, J. and Maggi, G. and Mahn, K.B.M. and Mancina, S. and Mandelartz, M. and Maruyama, R. and Mase, K. and Matis, H. and Maunu, R. and McNally, F. and McParland, C.P. and Meade, P. and Meagher, K. and Medici, M. and Meier, M. and Meli, A. and Menne, T. and Merino, G. and Meures, T. and Miarecki, S. and Minor, R.H. and Montaruli, T. and Moulai, M. and Murray, T. and Nahnhauer, R. and Naumann, U. and Neer, G. and Newcomb, M. and Niederhausen, H. and Nowicki, S.C. and Nygren, D.R. and Pollmann, A. Obertacke and Olivas, A. and O’Murchadha, A. and Palczewski, T. and Pandya, H. and Pankova, D.V. and Patton, S. and Peiffer, P. and Penek, Ö. and Pepper, J.A. and Heros, C. Pérez de los and Pettersen, C. and Pieloth, D. and Pinat, E. and Price, P.B. and Przybylski, G.T. and Quinnan, M. and Raab, C. and Rädel, L. and Rameez, M. and Rawlins, K. and Reimann, R. and Relethford, B. and Relich, M. and Resconi, E. and Rhode, W. and Richman, M. and Riedel, B. and Robertson, S. and Rongen, M. and Roucelle, C. and Rott, C. and Ruhe, T. and Ryckbosch, D. and Rysewyk, D. and Sabbatini, L. and Herrera, S.E. Sanchez and Sandrock, A. and Sandroos, J. and Sandstrom, P. and Sarkar, S. and Satalecka, K. and Schlunder, P. and Schmidt, T. and Schoenen, S. and Schöneberg, S. and Schukraft, A. and Schumacher, L. and Seckel, D. and Seunarine, S. and Solarz, M. and Soldin, D. and Song, M. and Spiczak, G.M. and Spiering, C. and Stanev, T. and Stasik, A. and Stettner, J. and Steuer, A. and Stezelberger, T. and Stokstad, R.G. and Stößl, A. and Ström, R. and Strotjohann, N.L. and Sulanke, K.-H. and Sullivan, G.W. and Sutherland, M. and Taavola, H. and Taboada, I. and Tatar, J. and Tenholt, F. and Ter-Antonyan, S. and Terliuk, A. and Tešić, G. and Thollander, L. and Tilav, S. and Toale, P.A. and Tobin, M.N. and Toscano, S. and Tosi, D. and Tselengidou, M. and Turcati, A. and Unger, E. and Usner, M. and Vandenbroucke, J. and Eijndhoven, N. van and Vanheule, S. and Rossem, M. van and Santen, J. van and Vehring, M. and Voge, M. and Vogel, E. and Vraeghe, M. and Wahl, D. and Walck, C. and Wallace, A. and Wallraff, M. and Wandkowsky, N. and Weaver, Ch. and Weiss, M.J. and Wendt, C. and Westerhoff, S. and Wharton, D. and Whelan, B.J. and Wickmann, S. and Wiebe, K. and Wiebusch, C.H. and Wille, L. and Williams, D.R. and Wills, L. and Wisniewski, P. and Wolf, M. and Wood, T.R. and Woolsey, E. and Woschnagg, K. and Xu, D.L. and Xu, X.W. and Xu, Y. and Yanez, J.P. and Yodh, G. and Yoshida, S. and Zoll, M.}, year={2017}, month=mar, pages={P03012} }

@ARTICLE{TXS2018,
       author = {{Aartsen}, M.~G. and {Ackermann}, M. and {Adams}, J. and {Aguilar}, J.~A. and {Ahlers}, M. and {Ahrens}, M. and {Al Samarai}, I. and {Altmann}, D. and {Andeen}, K. and {Anderson}, T. and {Ansseau}, I. and {Anton}, G. and {Arg{\"u}elles}, C. and {Auffenberg}, J. and {Axani}, S. and {Bagherpour}, H. and {Bai}, X. and {Barron}, J.~P. and {Barwick}, S.~W. and {Baum}, V. and {Bay}, R. and {Beatty}, J.~J. and {Becker Tjus}, J. and {Becker}, K.-H. and {BenZvi}, S. and {Berley}, D. and {Bernardini}, E. and {Besson}, D.~Z. and {Binder}, G. and {Bindig}, D. and {Blaufuss}, E. and {Blot}, S. and {Bohm}, C. and {B{\"o}rner}, M. and {Bos}, F. and {B{\"o}ser}, S. and {Botner}, O. and {Bourbeau}, E. and {Bourbeau}, J. and {Bradascio}, F. and {Braun}, J. and {Brenzke}, M. and {Bretz}, H.-P. and {Bron}, S. and {Brostean-Kaiser}, J. and {Burgman}, A. and {Busse}, R.~S. and {Carver}, T. and {Cheung}, E. and {Chirkin}, D. and {Christov}, A. and {Clark}, K. and {Classen}, L. and {Coenders}, S. and {Collin}, G.~H. and {Conrad}, J.~M. and {Coppin}, P. and {Correa}, P. and {Cowen}, D.~F. and {Cross}, R. and {Dave}, P. and {Day}, M. and {de Andr{\'e}}, J.~P.~A.~M. and {De Clercq}, C. and {DeLaunay}, J.~J. and {Dembinski}, H. and {De Ridder}, S. and {Desiati}, P. and {de Vries}, K.~D. and {de Wasseige}, G. and {de With}, M. and {DeYoung}, T. and {D{\'\i}az-V{\'e}lez}, J.~C. and {di Lorenzo}, V. and {Dujmovic}, H. and {Dumm}, J.~P. and {Dunkman}, M. and {Dvorak}, E. and {Eberhardt}, B. and {Ehrhardt}, T. and {Eichmann}, B. and {Eller}, P. and {Evenson}, P.~A. and {Fahey}, S. and {Fazely}, A.~R. and {Felde}, J. and {Filimonov}, K. and {Finley}, C. and {Flis}, S. and {Franckowiak}, A. and {Friedman}, E. and {Fritz}, A. and {Gaisser}, T.~K. and {Gallagher}, J. and {Gerhardt}, L. and {Ghorbani}, K. and {Glauch}, T. and {Gl{\"u}senkamp}, T. and {Goldschmidt}, A. and {Gonzalez}, J.~G. and {Grant}, D. and {Griffith}, Z. and {Haack}, C. and {Hallgren}, A. and {Halzen}, F. and {Hanson}, K. and {Hebecker}, D. and {Heereman}, D. and {Helbing}, K. and {Hellauer}, R. and {Hickford}, S. and {Hignight}, J. and {Hill}, G.~C. and {Hoffman}, K.~D. and {Hoffmann}, R. and {Hoinka}, T. and {Hokanson-Fasig}, B. and {Hoshina}, K. and {Huang}, F. and {Huber}, M. and {Hultqvist}, K. and {H{\"u}nnefeld}, M. and {Hussain}, R. and {In}, S. and {Iovine}, N. and {Ishihara}, A. and {Jacobi}, E. and {Japaridze}, G.~S. and {Jeong}, M. and {Jero}, K. and {Jones}, B.~J.~P. and {Kalaczynski}, P. and {Kang}, W. and {Kappes}, A. and {Kappesser}, D. and {Karg}, T. and {Karle}, A. and {Katz}, U. and {Kauer}, M. and {Keivani}, A. and {Kelley}, J.~L. and {Kheirandish}, A. and {Kim}, J. and {Kim}, M. and {Kintscher}, T. and {Kiryluk}, J. and {Kittler}, T. and {Klein}, S.~R. and {Koirala}, R. and {Kolanoski}, H. and {K{\"o}pke}, L. and {Kopper}, C. and {Kopper}, S. and {Koschinsky}, J.~P. and {Koskinen}, D.~J. and {Kowalski}, M. and {Krings}, K. and {Kroll}, M. and {Kr{\"u}ckl}, G. and {Kunwar}, S. and {Kurahashi}, N. and {Kuwabara}, T. and {Kyriacou}, A. and {Labare}, M. and {Lanfranchi}, J.~L. and {Larson}, M.~J. and {Lauber}, F. and {Leonard}, K. and {Lesiak-Bzdak}, M. and {Leuermann}, M. and {Liu}, Q.~R. and {Lozano Mariscal}, C.~J. and {Lu}, L. and {L{\"u}nemann}, J. and {Luszczak}, W. and {Madsen}, J. and {Maggi}, G. and {Mahn}, K.~B.~M. and {Mancina}, S. and {Maruyama}, R. and {Mase}, K. and {Maunu}, R. and {Meagher}, K. and {Medici}, M. and {Meier}, M. and {Menne}, T. and {Merino}, G. and {Meures}, T. and {Miarecki}, S. and {Micallef}, J. and {Moment{\'e}}, G. and {Montaruli}, T. and {Moore}, R.~W. and {Morse}, R. and {Moulai}, M. and {Nahnhauer}, R. and {Nakarmi}, P. and {Naumann}, U. and {Neer}, G.},
        title = "{Multimessenger observations of a flaring blazar coincident with high-energy neutrino IceCube-170922A}",
      journal = {Science},
         year = 2018,
        month = jul,
       volume = {361},
       number = {6398},
          eid = {eaat1378},
        pages = {eaat1378},
          doi = {10.1126/science.aat1378},
archivePrefix = {arXiv},
       eprint = {1807.08816},
 primaryClass = {astro-ph.HE},
       adsurl = {https://ui.adsabs.harvard.edu/abs/2018Sci...361.1378I}
}

@article{Aartsen:2016lmt,
    author = "Aartsen, M. G. and others",
    collaboration = "IceCube",
    title = "{The IceCube Realtime Alert System}",
    eprint = "1612.06028",
    archivePrefix = "arXiv",
    primaryClass = "astro-ph.HE",
    doi = "10.1016/j.astropartphys.2017.05.002",
    journal = "Astropart. Phys.",
    volume = "92",
    pages = "30--41",
    year = "2017"
}

@article{reusch2022candidate,
  title={Candidate tidal disruption event AT2019fdr coincident with a high-energy neutrino},
  author={Reusch, Simeon and Stein, Robert and Kowalski, Marek and Van Velzen, Sjoert and Franckowiak, Anna and Lunardini, Cecilia and Murase, Kohta and Winter, Walter and Miller-Jones, James CA and Kasliwal, Mansi M and others},
  journal={Physical review letters},
  volume={128},
  number={22},
  pages={221101},
  year={2022},
  publisher={APS}
}

@article{necker2023searchhighenergyneutrinostdelike,
      title={Search for High-Energy Neutrinos from TDE-like Flares with IceCube}, 
      author={Jannis Necker},
      year={2023},
      collaboration="IceCube",
      doi="10.22323/1.444.1478",
      journal="PoS",
      volume="ICRC2023",
      pages="1478",
}

@article{Necker_2025,
   author={Necker, J. and others},
   title={Flaires: A comprehensive catalog of dust echo-like infrared flares},
   journal={A\&A},
   volume={695},
   ISSN={1432-0746},
   url={http://dx.doi.org/10.1051/0004-6361/202451340},
   DOI={10.1051/0004-6361/202451340},
   publisher={EDP Sciences},
   year={2025},
   pages={A228} }

@article{van_Velzen_Stein_Gilfanov_Kowalski_Hayasaki_Reusch_Yao_Garrappa_Franckowiak_Gezari_et_al._2024, title={Establishing accretion flares from supermassive black holes as a source of high-energy neutrinos}, volume={529}, ISSN={0035-8711}, DOI={10.1093/mnras/stae610}, number={3}, journal={Monthly Notices of the Royal Astronomical Society}, author={van Velzen, Sjoert and Stein, Robert and Gilfanov, Marat and Kowalski, Marek and Hayasaki, Kimitake and Reusch, Simeon and Yao, Yuhan and Garrappa, Simone and Franckowiak, Anna and Gezari, Suvi and Nordin, Jakob and Fremling, Christoffer and Sharma, Yashvi and Yan, Lin and Kool, Erik C and Stern, Daniel and Veres, Patrik M and Sollerman, Jesper and Medvedev, Pavel and Sunyaev, Rashid and Bellm, Eric C and Dekany, Richard G and Duev, Dimitri A and Graham, Matthew J and Kasliwal, Mansi M and Kulkarni, Shrinivas R and Laher, Russ R and Riddle, Reed L and Rusholme, Ben}, year={2024}, month=apr, pages={2559–2576} }

@article{Braun_Dumm_De_Palma_Finley_Karle_Montaruli_2008, title={Methods for point source analysis in high energy neutrino telescopes}, volume={29}, rights={https://www.elsevier.com/tdm/userlicense/1.0/}, ISSN={09276505}, DOI={10.1016/j.astropartphys.2008.02.007}, number={4}, journal={Astroparticle Physics}, author={Braun, Jim and Dumm, Jon and De Palma, Francesco and Finley, Chad and Karle, Albrecht and Montaruli, Teresa}, year={2008}, month=may, pages={299–305}, language={en} }

@article{Ivezic_Kahn_Tyson_Abel_Acosta_Allsman_Alonso_AlSayyad_Anderson_Andrew_et_al._2019, title={LSST: From Science Drivers to Reference Design and Anticipated Data Products}, volume={873}, ISSN={0004-637X, 1538-4357}, DOI={10.3847/1538-4357/ab042c}, number={2}, journal={The Astrophysical Journal}, author={Ivezi{\'c}, {\v Z}eljko and Kahn, Steven M. and Tyson, J. Anthony and Abel, Bob and Acosta, Emily and Allsman, Robyn and Alonso, David and AlSayyad, Yusra and Anderson, Scott F. and Andrew, John and P. Angel, James Roger and Angeli, George Z. and Ansari, Reza and Antilogus, Pierre and Araujo, Constanza and Armstrong, Robert and Arndt, Kirk T. and Astier, Pierre and Aubourg, Éric and Auza, Nicole and Axelrod, Tim S. and Bard, Deborah J. and Barr, Jeff D. and Barrau, Aurelian and Bartlett, James G. and Bauer, Amanda E. and Bauman, Brian J. and Baumont, Sylvain and Bechtol, Ellen and Bechtol, Keith and Becker, Andrew C. and Becla, Jacek and Beldica, Cristina and Bellavia, Steve and Bianco, Federica B. and Biswas, Rahul and Blanc, Guillaume and Blazek, Jonathan and Blandford, Roger D. and Bloom, Josh S. and Bogart, Joanne and Bond, Tim W. and Booth, Michael T. and Borgland, Anders W. and Borne, Kirk and Bosch, James F. and Boutigny, Dominique and Brackett, Craig A. and Bradshaw, Andrew and Brandt, William Nielsen and Brown, Michael E. and Bullock, James S. and Burchat, Patricia and Burke, David L. and Cagnoli, Gianpietro and Calabrese, Daniel and Callahan, Shawn and Callen, Alice L. and Carlin, Jeffrey L. and Carlson, Erin L. and Chandrasekharan, Srinivasan and Charles-Emerson, Glenaver and Chesley, Steve and Cheu, Elliott C. and Chiang, Hsin-Fang and Chiang, James and Chirino, Carol and Chow, Derek and Ciardi, David R. and Claver, Charles F. and Cohen-Tanugi, Johann and Cockrum, Joseph J. and Coles, Rebecca and Connolly, Andrew J. and Cook, Kem H. and Cooray, Asantha and Covey, Kevin R. and Cribbs, Chris and Cui, Wei and Cutri, Roc and Daly, Philip N. and Daniel, Scott F. and Daruich, Felipe and Daubard, Guillaume and Daues, Greg and Dawson, William and Delgado, Francisco and Dellapenna, Alfred and Peyster, Robert De and Val-Borro, Miguel De and Digel, Seth W. and Doherty, Peter and Dubois, Richard and Dubois-Felsmann, Gregory P. and Durech, Josef and Economou, Frossie and Eifler, Tim and Eracleous, Michael and Emmons, Benjamin L. and Neto, Angelo Fausti and Ferguson, Henry and Figueroa, Enrique and Fisher-Levine, Merlin and Focke, Warren and Foss, Michael D. and Frank, James and Freemon, Michael D. and Gangler, Emmanuel and Gawiser, Eric and Geary, John C. and Gee, Perry and Geha, Marla and Gessner, Charles J. B. and Gibson, Robert R. and Gilmore, D. Kirk and Glanzman, Thomas and Glick, William and Goldina, Tatiana and Goldstein, Daniel A. and Goodenow, Iain and Graham, Melissa L. and Gressler, William J. and Gris, Philippe and Guy, Leanne P. and Guyonnet, Augustin and Haller, Gunther and Harris, Ron and Hascall, Patrick A. and Haupt, Justine and Hernandez, Fabio and Herrmann, Sven and Hileman, Edward and Hoblitt, Joshua and Hodgson, John A. and Hogan, Craig and Howard, James D. and Huang, Dajun and Huffer, Michael E. and Ingraham, Patrick and Innes, Walter R. and Jacoby, Suzanne H. and Jain, Bhuvnesh and Jammes, Fabrice and Jee, M. James and Jenness, Tim and Jernigan, Garrett and Jevremović, Darko and Johns, Kenneth and Johnson, Anthony S. and Johnson, Margaret W. G. and Jones, R. Lynne and Juramy-Gilles, Claire and Jurić, Mario and Kalirai, Jason S. and Kallivayalil, Nitya J. and Kalmbach, Bryce and Kantor, Jeffrey P. and Karst, Pierre and Kasliwal, Mansi M. and Kelly, Heather and Kessler, Richard and Kinnison, Veronica and Kirkby, David and Knox, Lloyd and Kotov, Ivan V. and Krabbendam, Victor L. and Krughoff, K. Simon and Kubánek, Petr and Kuczewski, John and Kulkarni, Shri and Ku, John and Kurita, Nadine R. and Lage, Craig S. and Lambert, Ron and Lange, Travis and Langton, J. Brian and Guillou, Laurent Le and Levine, Deborah and Liang, Ming and Lim, Kian-Tat and Lintott, Chris J. and Long, Kevin E. and Lopez, Margaux and Lotz, Paul J. and Lupton, Robert H. and Lust, Nate B. and MacArthur, Lauren A. and Mahabal, Ashish and Mandelbaum, Rachel and Markiewicz, Thomas W. and Marsh, Darren S. and Marshall, Philip J. and Marshall, Stuart and May, Morgan and McKercher, Robert and McQueen, Michelle and Meyers, Joshua and Migliore, Myriam and Miller, Michelle and Mills, David J. and Miraval, Connor and Moeyens, Joachim and Moolekamp, Fred E. and Monet, David G. and Moniez, Marc and Monkewitz, Serge and Montgomery, Christopher and Morrison, Christopher B. and Mueller, Fritz and Muller, Gary P. and Arancibia, Freddy Muñoz and Neill, Douglas R. and Newbry, Scott P. and Nief, Jean-Yves and Nomerotski, Andrei and Nordby, Martin and O’Connor, Paul and Oliver, John and Olivier, Scot S. and Olsen, Knut and O’Mullane, William and Ortiz, Sandra and Osier, Shawn and Owen, Russell E. and Pain, Reynald and Palecek, Paul E. and Parejko, John K. and Parsons, James B. and Pease, Nathan M. and Peterson, J. Matt and Peterson, John R. and Petravick, Donald L. and Petrick, M. E. Libby and Petry, Cathy E. and Pierfederici, Francesco and Pietrowicz, Stephen and Pike, Rob and Pinto, Philip A. and Plante, Raymond and Plate, Stephen and Plutchak, Joel P. and Price, Paul A. and Prouza, Michael and Radeka, Veljko and Rajagopal, Jayadev and Rasmussen, Andrew P. and Regnault, Nicolas and Reil, Kevin A. and Reiss, David J. and Reuter, Michael A. and Ridgway, Stephen T. and Riot, Vincent J. and Ritz, Steve and Robinson, Sean and Roby, William and Roodman, Aaron and Rosing, Wayne and Roucelle, Cecille and Rumore, Matthew R. and Russo, Stefano and Saha, Abhijit and Sassolas, Benoit and Schalk, Terry L. and Schellart, Pim and Schindler, Rafe H. and Schmidt, Samuel and Schneider, Donald P. and Schneider, Michael D. and Schoening, William and Schumacher, German and Schwamb, Megan E. and Sebag, Jacques and Selvy, Brian and Sembroski, Glenn H. and Seppala, Lynn G. and Serio, Andrew and Serrano, Eduardo and Shaw, Richard A. and Shipsey, Ian and Sick, Jonathan and Silvestri, Nicole and Slater, Colin T. and Smith, J. Allyn and Smith, R. Chris and Sobhani, Shahram and Soldahl, Christine and Storrie-Lombardi, Lisa and Stover, Edward and Strauss, Michael A. and Street, Rachel A. and Stubbs, Christopher W. and Sullivan, Ian S. and Sweeney, Donald and Swinbank, John D. and Szalay, Alexander and Takacs, Peter and Tether, Stephen A. and Thaler, Jon J. and Thayer, John Gregg and Thomas, Sandrine and Thornton, Adam J. and Thukral, Vaikunth and Tice, Jeffrey and Trilling, David E. and Turri, Max and Berg, Richard Van and Berk, Daniel Vanden and Vetter, Kurt and Virieux, Francoise and Vucina, Tomislav and Wahl, William and Walkowicz, Lucianne and Walsh, Brian and Walter, Christopher W. and Wang, Daniel L. and Wang, Shin-Yawn and Warner, Michael and Wiecha, Oliver and Willman, Beth and Winters, Scott E. and Wittman, David and Wolff, Sidney C. and Wood-Vasey, W. Michael and Wu, Xiuqin and Xin, Bo and Yoachim, Peter and Zhan, Hu}, year={2019}, month=mar, pages={111} }

@article{farrar2009giant,
  title={Giant AGN flares and cosmic ray bursts},
  author={Farrar, Glennys R and Gruzinov, Andrei},
  journal={The Astrophysical Journal},
  volume={693},
  number={1},
  pages={329--332},
  year={2009},
  publisher={The American Astronomical Society}
}

@article{Rees_1988, title={Tidal disruption of stars by black holes of 106–108 solar masses in nearby galaxies}, volume={333}, rights={http://www.springer.com/tdm}, ISSN={0028-0836, 1476-4687}, DOI={10.1038/333523a0}, number={6173}, journal={Nature}, author={Rees, Martin J.}, year={1988}, month=jun, pages={523–528}, language={en} }

@article{bellm2019zwicky,
  title={The Zwicky Transient Facility: system overview, performance, and first results},
  author={Bellm, Eric C and Kulkarni, Shrinivas R and Graham, Matthew J and Dekany, Richard and Smith, Roger M and Riddle, Reed and Masci, Frank J and Helou, George and Prince, Thomas A and Adams, Scott M and others},
  journal={Publications of the Astronomical Society of the Pacific},
  volume={131},
  number={995},
  pages={018002},
  year={2019},
  publisher={The Astronomical Society of the Pacific}
}

@article{Lu_Kumar_Evans_2016, title={Infrared emission from tidal disruption events – probing the pc-scale dust content around galactic nuclei}, volume={458}, ISSN={0035-8711, 1365-2966}, DOI={10.1093/mnras/stw307}, number={1}, journal={Monthly Notices of the Royal Astronomical Society}, author={Lu, Wenbin and Kumar, Pawan and Evans, Neal J.}, year={2016}, month=may, pages={575–581}, language={en} }

@ARTICLE{Komossa_Greiner_1999,
       author = {{Komossa}, Stefanie and {Greiner}, Jochen},
        title = "{Discovery of a giant and luminous X-ray outburst from the optically inactive galaxy pair RX J1242.6-1119}",
      journal = {\aap},
         year = 1999,
        month = sep,
       volume = {349},
        pages = {L45-L48},
          doi = {10.48550/arXiv.astro-ph/9908216},
archivePrefix = {arXiv},
       eprint = {astro-ph/9908216},
 primaryClass = {astro-ph},
       adsurl = {https://ui.adsabs.harvard.edu/abs/1999A&A...349L..45K}
}

@article{Gezari_2021, title={Tidal Disruption Events}, volume={59}, ISSN={0066-4146, 1545-4282}, DOI={10.1146/annurev-astro-111720-030029}, number={1}, journal={Annual Review of Astronomy and Astrophysics}, author={Gezari, Suvi}, year={2021}, month=sep, pages={21–58}, language={en} }

@article{Gezari_2006, title={Ultraviolet Detection of the Tidal Disruption of a Star by a Supermassive Black Hole}, volume={653}, ISSN={0004-637X, 1538-4357}, DOI={10.1086/509918}, number={1}, journal={The Astrophysical Journal}, author={Gezari, S. and Martin, D. C. and Milliard, B. and Basa, S. and Halpern, J. P. and Forster, K. and Friedman, P. G. and Morrissey, P. and Neff, S. G. and Schiminovich, D. and Seibert, M. and Small, T. and Wyder, T. K.}, year={2006}, month=dec, pages={L25–L28}, language={en} }

@article{masterson2024new,
  title={A new population of mid-infrared-selected tidal disruption events: Implications for tidal disruption event rates and host galaxy properties},
  author={Masterson, Megan and De, Kishalay and Panagiotou, Christos and Kara, Erin and Arcavi, Iair and Eilers, Anna-Christina and Frostig, Danielle and Gezari, Suvi and Grotova, Iuliia and Liu, Zhu and others},
  journal={The Astrophysical Journal},
  volume={961},
  number={2},
  pages={211},
  year={2024},
  publisher={IOP Publishing}
}

@article{sun2015extragalactic,
  title={Extragalactic high-energy transients: event rate densities and luminosity functions},
  author={Sun, Hui and Zhang, Bing and Li, Zhuo},
  journal={The Astrophysical Journal},
  volume={812},
  number={1},
  pages={33},
  year={2015},
  publisher={The American Astronomical Society}
}

@article{Aartsen_Ackermann_Adams_Aguilar_Ahlers_Ahrens_Alispach_Andeen_Anderson_Ansseau_et_al._2020, title={Time-Integrated Neutrino Source Searches with 10 Years of IceCube Data}, volume={124}, ISSN={0031-9007, 1079-7114}, DOI={10.1103/PhysRevLett.124.051103}, number={5}, journal={Physical Review Letters}, author={Aartsen, M. G. and Ackermann, M. and Adams, J. and Aguilar, J. A. and Ahlers, M. and Ahrens, M. and Alispach, C. and Andeen, K. and Anderson, T. and Ansseau, I. and Anton, G. and Argüelles, C. and Auffenberg, J. and Axani, S. and Backes, P. and Bagherpour, H. and Bai, X. and Balagopal, A. and Barbano, A. and Barwick, S. W. and Bastian, B. and Baum, V. and Baur, S. and Bay, R. and Beatty, J. J. and Becker, K.-H. and Becker Tjus, J. and BenZvi, S. and Berley, D. and Bernardini, E. and Besson, D. Z. and Binder, G. and Bindig, D. and Blaufuss, E. and Blot, S. and Bohm, C. and Börner, M. and Böser, S. and Botner, O. and Böttcher, J. and Bourbeau, E. and Bourbeau, J. and Bradascio, F. and Braun, J. and Bron, S. and Brostean-Kaiser, J. and Burgman, A. and Buscher, J. and Busse, R. S. and Carver, T. and Chen, C. and Cheung, E. and Chirkin, D. and Choi, S. and Clark, K. and Classen, L. and Coleman, A. and Collin, G. H. and Conrad, J. M. and Coppin, P. and Correa, P. and Cowen, D. F. and Cross, R. and Dave, P. and De Clercq, C. and DeLaunay, J. J. and Dembinski, H. and Deoskar, K. and De Ridder, S. and Desiati, P. and De Vries, K. D. and De Wasseige, G. and De With, M. and DeYoung, T. and Diaz, A. and D{\'i}az-V{\'e}lez, J. C. and Dujmovic, H. and Dunkman, M. and Dvorak, E. and Eberhardt, B. and Ehrhardt, T. and Eller, P. and Engel, R. and Evenson, P. A. and Fahey, S. and Fazely, A. R. and Felde, J. and Filimonov, K. and Finley, C. and Fox, D. and Franckowiak, A. and Friedman, E. and Fritz, A. and Gaisser, T. K. and Gallagher, J. and Ganster, E. and Garrappa, S. and Gerhardt, L. and Ghorbani, K. and Glauch, T. and Glüsenkamp, T. and Goldschmidt, A. and Gonzalez, J. G. and Grant, D. and Griffith, Z. and Griswold, S. and Günder, M. and Gündüz, M. and Haack, C. and Hallgren, A. and Halliday, R. and Halve, L. and Halzen, F. and Hanson, K. and Haungs, A. and Hebecker, D. and Heereman, D. and Heix, P. and Helbing, K. and Hellauer, R. and Henningsen, F. and Hickford, S. and Hignight, J. and Hill, G. C. and Hoffman, K. D. and Hoffmann, R. and Hoinka, T. and Hokanson-Fasig, B. and Hoshina, K. and Huang, F. and Huber, M. and Huber, T. and Hultqvist, K. and Hünnefeld, M. and Hussain, R. and In, S. and Iovine, N. and Ishihara, A. and Japaridze, G. S. and Jeong, M. and Jero, K. and Jones, B. J. P. and Jonske, F. and Joppe, R. and Kang, D. and Kang, W. and Kappes, A. and Kappesser, D. and Karg, T. and Karl, M. and Karle, A. and Katz, U. and Kauer, M. and Kelley, J. L. and Kheirandish, A. and Kim, J. and Kintscher, T. and Kiryluk, J. and Kittler, T. and Klein, S. R. and Koirala, R. and Kolanoski, H. and Köpke, L. and Kopper, C. and Kopper, S. and Koskinen, D. J. and Kowalski, M. and Krings, K. and Krückl, G. and Kulacz, N. and Kurahashi, N. and Kyriacou, A. and Labare, M. and Lanfranchi, J. L. and Larson, M. J. and Lauber, F. and Lazar, J. P. and Leonard, K. and Leszczyńska, A. and Leuermann, M. and Liu, Q. R. and Lohfink, E. and Lozano Mariscal, C. J. and Lu, L. and Lucarelli, F. and Lünemann, J. and Luszczak, W. and Lyu, Y. and Ma, W. Y. and Madsen, J. and Maggi, G. and Mahn, K. B. M. and Makino, Y. and Mallik, P. and Mallot, K. and Mancina, S. and Mariş, I. C. and Maruyama, R. and Mase, K. and Matis, H. S. and Maunu, R. and McNally, F. and Meagher, K. and Medici, M. and Medina, A. and Meier, M. and Meighen-Berger, S. and Menne, T. and Merino, G. and Meures, T. and Micallef, J. and Mockler, D. and Momenté, G. and Montaruli, T. and Moore, R. W. and Morse, R. and Moulai, M. and Muth, P. and Nagai, R. and Naumann, U. and Neer, G. and Niederhausen, H. and Nisa, M. U. and Nowicki, S. C. and Nygren, D. R. and Obertacke Pollmann, A. and Oehler, M. and Olivas, A. and O’Murchadha, A. and O’Sullivan, E. and Palczewski, T. and Pandya, H. and Pankova, D. V. and Park, N. and Peiffer, P. and Pérez De Los Heros, C. and Philippen, S. and Pieloth, D. and Pinat, E. and Pizzuto, A. and Plum, M. and Porcelli, A. and Price, P. B. and Przybylski, G. T. and Raab, C. and Raissi, A. and Rameez, M. and Rauch, L. and Rawlins, K. and Rea, I. C. and Reimann, R. and Relethford, B. and Renschler, M. and Renzi, G. and Resconi, E. and Rhode, W. and Richman, M. and Robertson, S. and Rongen, M. and Rott, C. and Ruhe, T. and Ryckbosch, D. and Rysewyk, D. and Safa, I. and Sanchez Herrera, S. E. and Sandrock, A. and Sandroos, J. and Santander, M. and Sarkar, S. and Sarkar, S. and Satalecka, K. and Schaufel, M. and Schieler, H. and Schlunder, P. and Schmidt, T. and Schneider, A. and Schneider, J. and Schröder, F. G. and Schumacher, L. and Sclafani, S. and Seckel, D. and Seunarine, S. and Shefali, S. and Silva, M. and Snihur, R. and Soedingrekso, J. and Soldin, D. and Song, M. and Spiczak, G. M. and Spiering, C. and Stachurska, J. and Stamatikos, M. and Stanev, T. and Stein, R. and Steinmüller, P. and Stettner, J. and Steuer, A. and Stezelberger, T. and Stokstad, R. G. and Stößl, A. and Strotjohann, N. L. and Stürwald, T. and Stuttard, T. and Sullivan, G. W. and Taboada, I. and Tenholt, F. and Ter-Antonyan, S. and Terliuk, A. and Tilav, S. and Tollefson, K. and Tomankova, L. and Tönnis, C. and Toscano, S. and Tosi, D. and Trettin, A. and Tselengidou, M. and Tung, C. F. and Turcati, A. and Turcotte, R. and Turley, C. F. and Ty, B. and Unger, E. and Unland Elorrieta, M. A. and Usner, M. and Vandenbroucke, J. and Van Driessche, W. and Van Eijk, D. and Van Eijndhoven, N. and Vanheule, S. and Van Santen, J. and Vraeghe, M. and Walck, C. and Wallace, A. and Wallraff, M. and Wandkowsky, N. and Watson, T. B. and Weaver, C. and Weindl, A. and Weiss, M. J. and Weldert, J. and Wendt, C. and Werthebach, J. and Whelan, B. J. and Whitehorn, N. and Wiebe, K. and Wiebusch, C. H. and Wille, L. and Williams, D. R. and Wills, L. and Wolf, M. and Wood, J. and Wood, T. R. and Woschnagg, K. and Wrede, G. and Xu, D. L. and Xu, X. W. and Xu, Y. and Yanez, J. P. and Yodh, G. and Yoshida, S. and Yuan, T. and Zöcklein, M.}, year={2020}, month=feb, pages={051103}, language={en} }

@inproceedings{Stein_IceCube_Collaboration_2019, address={Madison, WI, U.S.A.}, title={Search for Neutrinos from Populations of Optical Transients}, url={https://pos.sissa.it/358/1016}, DOI={10.22323/1.358.1016}, booktitle={Proceedings of 36th International Cosmic Ray Conference — PoS(ICRC2019)}, publisher={Sissa Medialab}, author= {Stein, Robert and others}, year={2019}, month=jul, pages={1016}, language={en} }

@ARTICLE{Mukhopadhyay_Wusinich_Murase_2026,
       author = {{Mukhopadhyay}, Mainak and {Wusinich}, Patrick and {Murase}, Kohta},
        title = "{Constraining high-energy neutrinos from tidal disruption events with IceCube high-energy starting events}",
      journal = {arXiv e-prints},
         year = 2026,
        month = jan,
          eid = {arXiv:2601.20934},
        pages = {arXiv:2601.20934},
          doi = {10.48550/arXiv.2601.20934},
archivePrefix = {arXiv},
       eprint = {2601.20934},
 primaryClass = {astro-ph.HE},
       adsurl = {https://ui.adsabs.harvard.edu/abs/2026arXiv260120934M}
}

@ARTICLE{Karmen_Gezari_Norman_Guolo_2026,
       author = {{Karmen}, Mitchell and {Gezari}, Suvi and {Norman}, Colin and {Guolo}, Muryel},
        title = "{Tidal Disruption Event Rates across Cosmic Time: Forecasts for LSST, Roman, and JWST and Their Constraints on the Supermassive Black Hole Mass Function}",
      journal = {\apj},
         year = 2026,
        month = jul,
       volume = {1006},
       number = {1},
          eid = {20},
        pages = {20},
          doi = {10.3847/1538-4357/ae7a49},
archivePrefix = {arXiv},
       eprint = {2602.04947},
 primaryClass = {astro-ph.HE},
       adsurl = {https://ui.adsabs.harvard.edu/abs/2026ApJ..1006...20K}
}

@article{Aartsen_Abbasi_Ackermann_Adams_Aguilar_Ahlers_Ahrens_Alispach_Allison_Amin_et_al._2021, title={IceCube-Gen2: the window to the extreme Universe}, volume={48}, ISSN={0954-3899, 1361-6471}, DOI={10.1088/1361-6471/abbd48}, number={6}, journal={Journal of Physics G: Nuclear and Particle Physics}, author={Aartsen, M G and Abbasi, R and Ackermann, M and Adams, J and Aguilar, J A and Ahlers, M and Ahrens, M and Alispach, C and Allison, P and Amin, N M and Andeen, K and Anderson, T and Ansseau, I and Anton, G and Argüelles, C and Arlen, T C and Auffenberg, J and Axani, S and Bagherpour, H and Bai, X and Balagopal V, A and Barbano, A and Bartos, I and Bastian, B and Basu, V and Baum, V and Baur, S and Bay, R and Beatty, J J and Becker, K-H and Tjus, J Becker and BenZvi, S and Berley, D and Bernardini, E and Besson, D Z and Binder, G and Bindig, D and Blaufuss, E and Blot, S and Bohm, C and Bohmer, M and Böser, S and Botner, O and Böttcher, J and Bourbeau, E and Bourbeau, J and Bradascio, F and Braun, J and Bron, S and Brostean-Kaiser, J and Burgman, A and Burley, R T and Buscher, J and Busse, R S and Bustamante, M and Campana, M A and Carnie-Bronca, E G and Carver, T and Chen, C and Chen, P and Cheung, E and Chirkin, D and Choi, S and Clark, B A and Clark, K and Classen, L and Coleman, A and Collin, G H and Connolly, A and Conrad, J M and Coppin, P and Correa, P and Cowen, D F and Cross, R and Dave, P and Deaconu, C and De Clercq, C and DeLaunay, J J and De Kockere, S and Dembinski, H and Deoskar, K and De Ridder, S and Desai, A and Desiati, P and De Vries, K D and De Wasseige, G and De With, M and DeYoung, T and Dharani, S and Diaz, A and D{\'i}az-V{\'e}lez, J C and Dujmovic, H and Dunkman, M and DuVernois, M A and Dvorak, E and Ehrhardt, T and Eller, P and Engel, R and Evans, J J and Evenson, P A and Fahey, S and Farrag, K and Fazely, A R and Felde, J and Fienberg, A T and Filimonov, K and Finley, C and Fischer, L and Fox, D and Franckowiak, A and Friedman, E and Fritz, A and Gaisser, T K and Gallagher, J and Ganster, E and Garcia-Fernandez, D and Garrappa, S and Gartner, A and Gerhard, L and Gernhaeuser, R and Ghadimi, A and Glaser, C and Glauch, T and Glüsenkamp, T and Goldschmidt, A and Gonzalez, J G and Goswami, S and Grant, D and Grégoire, T and Griffith, Z and Griswold, S and Gündüz, M and Haack, C and Hallgren, A and Halliday, R and Halve, L and Halzen, F and Hanson, J C and Hanson, K and Hardin, J and Haugen, J and Haungs, A and Hauser, S and Hebecker, D and Heinen, D and Heix, P and Helbing, K and Hellauer, R and Henningsen, F and Hickford, S and Hignight, J and Hill, C and Hill, G C and Hoffman, K D and Hoffmann, B and Hoffmann, R and Hoinka, T and Hokanson-Fasig, B and Holzapfel, K and Hoshina, K and Huang, F and Huber, M and Huber, T and Huege, T and Hughes, K and Hultqvist, K and Hünnefeld, M and Hussain, R and In, S and Iovine, N and Ishihara, A and Jansson, M and Japaridze, G S and Jeong, M and Jones, B J P and Jonske, F and Joppe, R and Kalekin, O and Kang, D and Kang, W and Kang, X and Kappes, A and Kappesser, D and Karg, T and Karl, M and Karle, A and Katori, T and Katz, U and Kauer, M and Keivani, A and Kellermann, M and Kelley, J L and Kheirandish, A and Kim, J and Kin, K and Kintscher, T and Kiryluk, J and Kittler, T and Kleifges, M and Klein, S R and Koirala, R and Kolanoski, H and Köpke, L and Kopper, C and Kopper, S and Koskinen, D J and Koundal, P and Kovacevich, M and Kowalski, M and Krauss, C B and Krings, K and Krückl, G and Kulacz, N and Kurahashi, N and Gualda, C Lagunas and Lahmann, R and Lanfranchi, J L and Larson, M J and Latif, U and Lauber, F and Lazar, J P and Leonard, K and Leszczyńska, A and Li, Y and Liu, Q R and Lohfink, E and LoSecco, J and Mariscal, C J Lozano and Lu, L and Lucarelli, F and Ludwig, A and Lünemann, J and Luszczak, W and Lyu, Y and Ma, W Y and Madsen, J and Maggi, G and Mahn, K B M and Makino, Y and Mallik, P and Mancina, S and Mandalia, S and Mariş, I C and Marka, S and Marka, Z and Maruyama, R and Mase, K and Maunu, R and McNally, F and Meagher, K and Medina, A and Meier, M and Meighen-Berger, S and Merz, J and Meyers, Z S and Micallef, J and Mockler, D and Momenté, G and Montaruli, T and Moore, R W and Morse, R and Moulai, M and Muth, P and Naab, R and Nagai, R and Nam, J and Nauman, U and Necker, J and Neer, G and Nelles, A and Nguyễn, L V and Niederhausen, H and Nisa, M U and Nowicki, S C and Nygren, D R and Oberla, E and Pollmann, A Obertacke and Oehler, M and Olivas, A and O’Sullivan, E and Pan, Y and Pandya, H and Pankova, D V and Papp, L and Park, N and Parker, G K and Paudel, E N and Peiffer, P and Pérez De Los Heros, C and Petersen, T C and Philippen, S and Pieloth, D and Pieper, S and Pinfold, J L and Pizzuto, A and Plaisier, I and Plum, M and Popovych, Y and Porcelli, A and Rodriguez, M Prado and Price, P B and Przybylski, G T and Raab, C and Raissi, A and Rameez, M and Rauch, L and Rawlins, K and Rea, I C and Rehman, A and Reimann, R and Renschler, M and Renzi, G and Resconi, E and Reusch, S and Rhode, W and Richman, M and Riedel, B and Riegel, M and Roberts, E J and Robertson, S and Roellinghoff, G and Rongen, M and Rott, C and Ruhe, T and Ryckbosch, D and Cantu, D Rysewyk and Safa, I and Herrera, S E Sanchez and Sandrock, A and Sandroos, J and Sandstrom, P and Santander, M and Sarkar, S and Sarkar, S and Satalecka, K and Scharf, M and Schaufel, M and Schieler, H and Schlunder, P and Schmidt, T and Schneider, A and Schneider, J and Schröder, F G and Schumacher, L and Sclafani, S and Seckel, D and Seunarine, S and Shaevitz, M H and Sharma, A and Shefali, S and Silva, M and Smith, D and Smithers, B and Snihur, R and Soedingrekso, J and Soldin, D and Söldner-Rembold, S and Song, M and Southall, D and Spiczak, G M and Spiering, C and Stachurska, J and Stamatikos, M and Stanev, T and Stein, R and Stettner, J and Steuer, A and Stezelberger, T and Stokstad, R G and Strotjohann, N L and Stürwald, T and Stuttard, T and Sullivan, G W and Taboada, I and Taketa, A and Tanaka, H K M and Tenholt, F and Ter-Antonyan, S and Terliuk, A and Tilav, S and Tollefson, K and Tomankova, L and Tönnis, C and Torres, J and Toscano, S and Tosi, D and Trettin, A and Tselengidou, M and Tung, C F and Turcati, A and Turcotte, R and Turley, C F and Twagirayezu, J P and Ty, B and Unger, E and Elorrieta, M A Unland and Vandenbroucke, J and Van Eijk, D and Van Eijndhoven, N and Vannerom, D and Van Santen, J and Veberic, D and Verpoest, S and Vieregg, A and Vraeghe, M and Walck, C and Watson, T B and Weaver, C and Weindl, A and Weinstock, L and Weiss, M J and Weldert, J and Welling, C and Wendt, C and Werthebach, J and Whitehorn, N and Wiebe, K and Wiebusch, C H and Williams, D R and Wissel, S A and Wolf, M and Wood, T R and Woschnagg, K and Wrede, G and Wren, S and Wulff, J and Xu, X W and Xu, Y and Yanez, J P and Yoshida, S and Yuan, T and Zhang, Z and Zierke, S and Zöcklein, M}, year={2021}, month=jun, pages={060501} }

@article{Yao_Ravi_Gezari_Van_Velzen_Lu_Schulze_Somalwar_Kulkarni_Hammerstein_Nicholl_et_al._2023, title={Tidal Disruption Event Demographics with the Zwicky Transient Facility: Volumetric Rates, Luminosity Function, and Implications for the Local Black Hole Mass Function}, volume={955}, ISSN={2041-8205, 2041-8213}, DOI={10.3847/2041-8213/acf216}, number={1}, journal={The Astrophysical Journal Letters}, author={Yao, Yuhan and Ravi, Vikram and Gezari, Suvi and Van Velzen, Sjoert and Lu, Wenbin and Schulze, Steve and Somalwar, Jean J. and Kulkarni, S. R. and Hammerstein, Erica and Nicholl, Matt and Graham, Matthew J. and Perley, Daniel A. and Cenko, S. Bradley and Stein, Robert and Ricarte, Angelo and Chadayammuri, Urmila and Quataert, Eliot and Bellm, Eric C. and Bloom, Joshua S. and Dekany, Richard and Drake, Andrew J. and Groom, Steven L. and Mahabal, Ashish A. and Prince, Thomas A. and Riddle, Reed and Rusholme, Ben and Sharma, Yashvi and Sollerman, Jesper and Yan, Lin}, year={2023}, month=sep, pages={L6} }

@article{Wang_Liu_Dai_Cheng_2011, title={Probing the tidal disruption flares of massive black holes with high-energy neutrinos}, volume={84}, rights={http://link.aps.org/licenses/aps-default-license}, ISSN={1550-7998, 1550-2368}, DOI={10.1103/PhysRevD.84.081301}, number={8}, journal={Physical Review D}, author={Wang, Xiang-Yu and Liu, Ruo-Yu and Dai, Zi-Gao and Cheng, K. S.}, year={2011}, month=oct, pages={081301}, language={en} }

@ARTICLE{Evans_Kochanek_1989,
       author = {{Evans}, Charles R. and {Kochanek}, Christopher S.},
        title = "{The Tidal Disruption of a Star by a Massive Black Hole}",
      journal = {\apjl},
         year = 1989,
        month = nov,
       volume = {346},
        pages = {L13},
          doi = {10.1086/185567},
       adsurl = {https://ui.adsabs.harvard.edu/abs/1989ApJ...346L..13E}
}

@ARTICLE{van_Velzen_Mendez_Krolik_Gorjian_2016,
       author = {{van Velzen}, S. and {Mendez}, A.~J. and {Krolik}, J.~H. and {Gorjian}, V.},
        title = "{Discovery of Transient Infrared Emission from Dust Heated by Stellar Tidal Disruption Flares}",
      journal = {\apj},
         year = 2016,
        month = sep,
       volume = {829},
       number = {1},
          eid = {19},
        pages = {19},
          doi = {10.3847/0004-637X/829/1/19},
archivePrefix = {arXiv},
       eprint = {1605.04304},
 primaryClass = {astro-ph.HE},
       adsurl = {https://ui.adsabs.harvard.edu/abs/2016ApJ...829...19V}
}

@ARTICLE{Bloom_Giannios_Metzger_Cenko_Perley_Butler_Tanvir_Levan_Brien_Strubbe_et_al._2011,
       author = {{Bloom}, Joshua S. and {Giannios}, Dimitrios and {Metzger}, Brian D. and {Cenko}, S. Bradley and {Perley}, Daniel A. and {Butler}, Nathaniel R. and {Tanvir}, Nial R. and {Levan}, Andrew J. and {O'Brien}, Paul T. and {Strubbe}, Linda E. and {De Colle}, Fabio and {Ramirez-Ruiz}, Enrico and {Lee}, William H. and {Nayakshin}, Sergei and {Quataert}, Eliot and {King}, Andrew R. and {Cucchiara}, Antonino and {Guillochon}, James and {Bower}, Geoffrey C. and {Fruchter}, Andrew S. and {Morgan}, Adam N. and {van der Horst}, Alexander J.},
        title = "{A Possible Relativistic Jetted Outburst from a Massive Black Hole Fed by a Tidally Disrupted Star}",
      journal = {Science},
         year = 2011,
        month = jul,
       volume = {333},
       number = {6039},
        pages = {203},
          doi = {10.1126/science.1207150},
archivePrefix = {arXiv},
       eprint = {1104.3257},
 primaryClass = {astro-ph.HE},
       adsurl = {https://ui.adsabs.harvard.edu/abs/2011Sci...333..203B}
}

@ARTICLE{Burrows_Kennea_Ghisellini_Mangano_Zhang_Page_Eracleous_Romano_Sakamoto_Falcone_et_al._2011,
       author = {{Burrows}, D.~N. and {Kennea}, J.~A. and {Ghisellini}, G. and {Mangano}, V. and {Zhang}, B. and {Page}, K.~L. and {Eracleous}, M. and {Romano}, P. and {Sakamoto}, T. and {Falcone}, A.~D. and {Osborne}, J.~P. and {Campana}, S. and {Beardmore}, A.~P. and {Breeveld}, A.~A. and {Chester}, M.~M. and {Corbet}, R. and {Covino}, S. and {Cummings}, J.~R. and {D'Avanzo}, P. and {D'Elia}, V. and {Esposito}, P. and {Evans}, P.~A. and {Fugazza}, D. and {Gelbord}, J.~M. and {Hiroi}, K. and {Holland}, S.~T. and {Huang}, K.~Y. and {Im}, M. and {Israel}, G. and {Jeon}, Y. and {Jeon}, Y.-B. and {Jun}, H.~D. and {Kawai}, N. and {Kim}, J.~H. and {Krimm}, H.~A. and {Marshall}, F.~E. and {P. M{\'e}sz{\'a}ros} and {Negoro}, H. and {Omodei}, N. and {Park}, W.-K. and {Perkins}, J.~S. and {Sugizaki}, M. and {Sung}, H.-I. and {Tagliaferri}, G. and {Troja}, E. and {Ueda}, Y. and {Urata}, Y. and {Usui}, R. and {Antonelli}, L.~A. and {Barthelmy}, S.~D. and {Cusumano}, G. and {Giommi}, P. and {Melandri}, A. and {Perri}, M. and {Racusin}, J.~L. and {Sbarufatti}, B. and {Siegel}, M.~H. and {Gehrels}, N.},
        title = "{Relativistic jet activity from the tidal disruption of a star by a massive black hole}",
      journal = {\nat},
         year = 2011,
        month = aug,
       volume = {476},
       number = {7361},
        pages = {421-424},
          doi = {10.1038/nature10374},
archivePrefix = {arXiv},
       eprint = {1104.4787},
 primaryClass = {astro-ph.HE},
       adsurl = {https://ui.adsabs.harvard.edu/abs/2011Natur.476..421B}
}

@article{Jiang_Dou_Wang_Yang_Lyu_Zhou_2016, title={THE WISE DETECTION OF AN INFRARED ECHO IN TIDAL DISRUPTION EVENT ASASSN-14li}, volume={828}, ISSN={2041-8205, 2041-8213}, DOI={10.3847/2041-8205/828/1/L14}, number={1}, journal={The Astrophysical Journal Letters}, author={Jiang, Ning and Dou, Liming and Wang, Tinggui and Yang, Chenwei and Lyu, Jianwei and Zhou, Hongyan}, year={2016}, month=sep, pages={L14} }

@article{Stein_Reusch_Franckowiak_Kowalski_Necker_Weimann_Kasliwal_Sollerman_Ahumada_Amaro_Seoane_et_al._2023, title={Neutrino follow-up with the Zwicky transient facility: results from the first 24 campaigns}, volume={521}, rights={https://academic.oup.com/journals/pages/open_access/funder_policies/chorus/standard_publication_model}, ISSN={0035-8711, 1365-2966}, DOI={10.1093/mnras/stad767}, number={4}, journal={Monthly Notices of the Royal Astronomical Society}, author={Stein, Robert and Reusch, Simeon and Franckowiak, Anna and Kowalski, Marek and Necker, Jannis and Weimann, Sven and Kasliwal, Mansi M and Sollerman, Jesper and Ahumada, Tomas and Amaro Seoane, Pau and Anand, Shreya and Andreoni, Igor and Bellm, Eric C and Bloom, Joshua S and Coughlin, Michael and De, Kishalay and Fremling, Christoffer and Gezari, Suvi and Graham, Matthew and Groom, Steven L and Helou, George and Kaplan, David L and Karambelkar, Viraj and Kong, Albert K H and Kool, Erik C and Lincetto, Massimiliano and Mahabal, Ashish A and Masci, Frank J and Medford, Michael S and Morgan, Robert and Nordin, Jakob and Rodriguez, Hector and Sharma, Yashvi and van Santen, Jakob and van Velzen, Sjoert and Yan, Lin}, year={2023}, month=mar, pages={5046–5063}, language={en} }

@article{Necker_de_Jaeger_Stein_Franckowiak_Shappee_Kowalski_Kochanek_Stanek_Beacom_Desai_et_al._2022, title={ASAS-SN follow-up of IceCube high-energy neutrino alerts}, volume={516}, rights={https://academic.oup.com/journals/pages/open_access/funder_policies/chorus/standard_publication_model}, ISSN={0035-8711, 1365-2966}, DOI={10.1093/mnras/stac2261}, number={2}, journal={Monthly Notices of the Royal Astronomical Society}, author={Necker, Jannis and de Jaeger, Thomas and Stein, Robert and Franckowiak, Anna and Shappee, Benjamin J and Kowalski, Marek and Kochanek, Christopher S and Stanek, Krzysztof Z and Beacom, John F and Desai, Dhvanil D and Neumann, Kyle and Jayasinghe, Tharindu and Holoien, T W-S and Thompson, Todd A and Holmbo, Simon}, year={2022}, month=sep, pages={2455–2469}, language={en} }

@ARTICLE{icecube2022evidence,
       author = {{Abbasi}, R. and {Ackermann}, M. and {Adams}, J. and {Aguilar}, J.~A. and {Ahlers}, M. and {Ahrens}, M. and {Alameddine}, J.~M. and {Alispach}, C. and {Alves}, Jr., A.~A. and {Amin}, N.~M. and {Andeen}, K. and {Anderson}, T. and {Anton}, G. and {Arg{\"u}elles}, C. and {Ashida}, Y. and {Axani}, S. and {Bai}, X. and {Balagopal}, A.~V. and {Barbano}, V.~A. and {Barwick}, S.~W. and {Bastian}, B. and {Basu}, V. and {Baur}, S. and {Bay}, R. and {Beatty}, J.~J. and {Becker}, K.-H. and {Becker Tjus}, J. and {Bellenghi}, C. and {Benzvi}, S. and {Berley}, D. and {Bernardini}, E. and {Besson}, D.~Z. and {Binder}, G. and {Bindig}, D. and {Blaufuss}, E. and {Blot}, S. and {Boddenberg}, M. and {Bontempo}, F. and {Borowka}, J. and {B{\"o}ser}, S. and {Botner}, O. and {B{\"o}ttcher}, J. and {Bourbeau}, E. and {Bradascio}, F. and {Braun}, J. and {Brinson}, B. and {Bron}, S. and {Brostean-Kaiser}, J. and {Browne}, S. and {Burgman}, A. and {Burley}, R.~T. and {Busse}, R.~S. and {Campana}, M.~A. and {Carnie-Bronca}, E.~G. and {Chen}, C. and {Chen}, Z. and {Chirkin}, D. and {Choi}, K. and {Clark}, B.~A. and {Clark}, K. and {Classen}, L. and {Coleman}, A. and {Collin}, G.~H. and {Conrad}, J.~M. and {Coppin}, P. and {Correa}, P. and {Cowen}, D.~F. and {Cross}, R. and {Dappen}, C. and {Dave}, P. and {de Clercq}, C. and {Delaunay}, J.~J. and {Delgado L{\'o}pez}, D. and {Dembinski}, H. and {Deoskar}, K. and {Desai}, A. and {Desiati}, P. and {de Vries}, K.~D. and {de Wasseige}, G. and {de With}, M. and {Deyoung}, T. and {Diaz}, A. and {D{\'\i}az-V{\'e}lez}, J.~C. and {Dittmer}, M. and {Dujmovic}, H. and {Dunkman}, M. and {Duvernois}, M.~A. and {Dvorak}, E. and {Ehrhardt}, T. and {Eller}, P. and {Engel}, R. and {Erpenbeck}, H. and {Evans}, J. and {Evenson}, P.~A. and {Fan}, K.~L. and {Fazely}, A.~R. and {Fedynitch}, A. and {Feigl}, N. and {Fiedlschuster}, S. and {Fienberg}, A.~T. and {Filimonov}, K. and {Finley}, C. and {Fischer}, L. and {Fox}, D. and {Franckowiak}, A. and {Friedman}, E. and {Fritz}, A. and {F{\"u}rst}, P. and {Gaisser}, T.~K. and {Gallagher}, J. and {Ganster}, E. and {Garcia}, A. and {Garrappa}, S. and {Gerhardt}, L. and {Ghadimi}, A. and {Glaser}, C. and {Glauch}, T. and {Gl{\"u}senkamp}, T. and {Goldschmidt}, A. and {Gonzalez}, J.~G. and {Goswami}, S. and {Grant}, D. and {Gr{\'e}goire}, T. and {Griswold}, S. and {G{\"u}nther}, C. and {Gutjahr}, P. and {Haack}, C. and {Hallgren}, A. and {Halliday}, R. and {Halve}, L. and {Halzen}, F. and {Hanson}, M. Ha Minh K. and {Hardin}, J. and {Harnisch}, A.~A. and {Haungs}, A. and {Hebecker}, D. and {Helbing}, K. and {Henningsen}, F. and {Hettinger}, E.~C. and {Hickford}, S. and {Hignight}, J. and {Hill}, C. and {Hill}, G.~C. and {Hoffman}, K.~D. and {Hoffmann}, R. and {Hokanson-Fasig}, B. and {Hoshina}, K. and {Huang}, F. and {Huber}, M. and {Huber}, T. and {Hultqvist}, K. and {H{\"u}nnefeld}, M. and {Hussain}, R. and {Hymon}, K. and {in}, S. and {Iovine}, N. and {Ishihara}, A. and {Jansson}, M. and {Japaridze}, G.~S. and {Jeong}, M. and {Jin}, M. and {Jones}, B.~J.~P. and {Kang}, D. and {Kang}, W. and {Kang}, X. and {Kappes}, A. and {Kappesser}, D. and {Kardum}, L. and {Karg}, T. and {Karl}, M. and {Karle}, A. and {Katz}, U. and {Kauer}, M. and {Kellermann}, M. and {Kelley}, J.~L. and {Kheirandish}, A. and {Kin}, K. and {Kintscher}, T. and {Kiryluk}, J. and {Klein}, S.~R. and {Koirala}, R. and {Kolanoski}, H. and {Kontrimas}, T. and {K{\"o}pke}, L. and {Kopper}, C. and {Kopper}, S. and {Koskinen}, D.~J. and {Koundal}, P. and {Kovacevich}, M. and {Kowalski}, M. and {Kozynets}, T. and {Kun}, E. and {Kurahashi}, N. and {Lad}, N. and {Lagunas Gualda}, C. and {Lanfranchi}, J.~L. and {Larson}, M.~J. and {Lauber}, F. and {Lazar}, J.~P.},
        title = "{Evidence for neutrino emission from the nearby active galaxy NGC 1068}",
      journal = {Science},
         year = 2022,
        month = nov,
       volume = {378},
       number = {6619},
        pages = {538-543},
          doi = {10.1126/science.abg3395},
archivePrefix = {arXiv},
       eprint = {2211.09972},
 primaryClass = {astro-ph.HE},
       adsurl = {https://ui.adsabs.harvard.edu/abs/2022Sci...378..538I}
}

@article{bartos2025bayesian,
  title={Bayesian Deep-stacking for high-energy neutrino searches},
  author={Bartos, Imre and Ackermann, Markus and Kowalski, Marek},
  journal={Journal of Cosmology and Astroparticle Physics},
  volume={2025},
  number={06},
  pages={064},
  year={2025},
  publisher={IOP Publishing}
}

@article{IceCube_Collaboration_2013, author = {Aartsen, M. G. and Abbasi, R. and Abdou, Y. and others}, title={Evidence for High-Energy Extraterrestrial Neutrinos at the IceCube Detector}, volume={342}, ISSN={0036-8075, 1095-9203}, DOI={10.1126/science.1242856}, number={6161}, journal={Science}, year={2013}, month=nov, pages={1242856}, language={en} }

@article{Abbasi_Ackermann_Adams_Agarwalla_Aguilar_Ahlers_Alameddine_Ali_Amin_Andeen_et_al._2026, title={Evidence for Neutrino Emission from X-Ray-bright Active Galactic Nuclei with IceCube}, volume={1000}, ISSN={2041-8205, 2041-8213}, DOI={10.3847/2041-8213/ae4aad}, number={1}, journal={The Astrophysical Journal Letters}, author={Abbasi, R. and Ackermann, M. and Adams, J. and Agarwalla, S. K. and Aguilar, J. A. and Ahlers, M. and Alameddine, J. M. and Ali, S. and Amin, N. M. and Andeen, K. and Argüelles, C. and Ashida, Y. and Athanasiadou, S. and Axani, S. N. and Babu, R. and Bai, X. and Baines-Holmes, J. and Balagopal V., A. and Barwick, S. W. and Bash, S. and Basu, V. and Bay, R. and Beatty, J. J. and Becker Tjus, J. and Behrens, P. and Beise, J. and Bellenghi, C. and Benkel, B. and BenZvi, S. and Berley, D. and Bernardini, E. and Besson, D. Z. and Blaufuss, E. and Bloom, L. and Blot, S. and Bodo, I. and Bontempo, F. and Motzkin, J. Y. Book and Boscolo Meneguolo, C. and Böser, S. and Botner, O. and Böttcher, J. and Braun, J. and Brinson, B. and Brisson-Tsavoussis, Z. and Burley, R. T. and Butterfield, D. and Campana, M. A. and Carloni, K. and Carpio, J. and Chattopadhyay, S. and Chau, N. and Chen, Z. and Chirkin, D. and Choi, S. and Clark, B. A. and Coleman, A. and Coleman, P. and Collin, G. H. and Borja, D. A. Coloma and Connolly, A. and Conrad, J. M. and Cowen, D. F. and De Clercq, C. and DeLaunay, J. J. and Delgado, D. and Delmeulle, T. and Deng, S. and Desiati, P. and De Vries, K. D. and De Wasseige, G. and DeYoung, T. and D{\'i}az-V{\'e}lez, J. C. and DiKerby, S. and Ding, T. and Dittmer, M. and Domi, A. and Draper, L. and Dueser, L. and Durnford, D. and Dutta, K. and DuVernois, M. A. and Ehrhardt, T. and Eidenschink, L. and Eimer, A. and Eller, P. and Ellinger, E. and Elsässer, D. and Engel, R. and Erpenbeck, H. and Esmail, W. and Eulig, S. and Evans, J. and Evenson, P. A. and Fan, K. L. and Fang, K. and Farrag, K. and Fazely, A. R. and Fedynitch, A. and Feigl, N. and Finley, C. and Fischer, L. and Fox, D. and Franckowiak, A. and Fukami, S. and Fürst, P. and Gallagher, J. and Ganster, E. and Garcia, A. and Garcia, M. and Garg, G. and Genton, E. and Gerhardt, L. and Ghadimi, A. and Glüsenkamp, T. and Gonzalez, J. G. and Goswami, S. and Granados, A. and Grant, D. and Gray, S. J. and Griffin, S. and Griswold, S. and Groth, K. M. and Guevel, D. and Günther, C. and Gutjahr, P. and Ha, C. and Haack, C. and Hallgren, A. and Halve, L. and Halzen, F. and Hamacher, L. and Ha Minh, M. and Handt, M. and Hanson, K. and Hardin, J. and Harnisch, A. A. and Hatch, P. and Haungs, A. and Häussler, J. and Helbing, K. and Hellrung, J. and Henke, B. and Hennig, L. and Henningsen, F. and Heuermann, L. and Hewett, R. and Heyer, N. and Hickford, S. and Hidvegi, A. and Hill, C. and Hill, G. C. and Hmaid, R. and Hoffman, K. D. and Hooper, D. and Hori, S. and Hoshina, K. and Hostert, M. and Hou, W. and Hrywniak, M. and Huber, T. and Hultqvist, K. and Hymon, K. and Ishihara, A. and Iwakiri, W. and Jacquart, M. and Jain, S. and Janik, O. and Jansson, M. and Jeong, M. and Jin, M. and Kamp, N. and Kang, D. and Kang, W. and Kang, X. and Kappes, A. and Kardum, L. and Karg, T. and Karl, M. and Karle, A. and Katil, A. and Kauer, M. and Kelley, J. L. and Khanal, M. and Khatee Zathul, A. and Kheirandish, A. and Kimku, H. and Kiryluk, J. and Klein, C. and Klein, S. R. and Kobayashi, Y. and Kochocki, A. and Koirala, R. and Kolanoski, H. and Kontrimas, T. and Köpke, L. and Kopper, C. and Koskinen, D. J. and Koundal, P. and Kowalski, M. and Kozynets, T. and Kravka, A. and Krieger, N. and Krishnamoorthi, J. and Krishnan, T. and Kruiswijk, K. and Krupczak, E. and Kumar, A. and Kun, E. and Kurahashi, N. and Lad, N. and Lagunas Gualda, C. and Arnaud, L. Lallement and Lamoureux, M. and Larson, M. J. and Lauber, F. and Lazar, J. P. and Leonard DeHolton, K. and Leszczyńska, A. and Liao, J. and Lin, C. and Liu, Q. R. and Liu, Y. T. and Liubarska, M. and Love, C. and Lu, L. and Lucarelli, F. and Luszczak, W. and Lyu, Y. and Macdonald, M. and Madsen, J. and Magnus, E. and Makino, Y. and Manao, E. and Mancina, S. and Mand, A. and Mariş, I. C. and Marka, S. and Marka, Z. and Marten, L. and Martinez-Soler, I. and Maruyama, R. and Mauro, J. and Mayhew, F. and McNally, F. and Mead, J. V. and Meagher, K. and Mechbal, S. and Medina, A. and Meier, M. and Merckx, Y. and Merten, L. and Mitchell, J. and Molchany, L. and Mondal, S. and Montaruli, T. and Moore, R. W. and Morii, Y. and Mosbrugger, A. and Moulai, M. and Mousadi, D. and Moyaux, E. and Mukherjee, T. and Naab, R. and Nakos, M. and Naumann, U. and Necker, J. and Neste, L. and Neumann, M. and Niederhausen, H. and Nisa, M. U. and Noda, K. and Noell, A. and Novikov, A. and Obertacke, A. and O’Dell, V. and Olivas, A. and Orsoe, R. and Osborn, J. and O’Sullivan, E. and Palusova, V. and Pandya, H. and Parenti, A. and Park, N. and Parrish, V. and Paudel, E. N. and Paul, L. and De Los Heros, C. Pérez and Pernice, T. and Petersen, T. C. and Peterson, J. and Plum, M. and Pontén, A. and Poojyam, V. and Popovych, Y. and Rodriguez, M. Prado and Pries, B. and Procter-Murphy, R. and Przybylski, G. T. and Pyras, L. and Raab, C. and Rack-Helleis, J. and Rad, N. and Ravn, M. and Rawlins, K. and Rechav, Z. and Rehman, A. and Reistroffer, I. and Resconi, E. and Reusch, S. and Rho, C. D. and Rhode, W. and Ricca, L. and Riedel, B. and Rifaie, A. and Roberts, E. J. and Rongen, M. and Rosted, A. and Rott, C. and Ruhe, T. and Ruohan, L. and Ryckbosch, D. and Saffer, J. and Salazar-Gallegos, D. and Sampathkumar, P. and Sandrock, A. and Sanger-Johnson, G. and Santander, M. and Sarkar, S. and Savelberg, J. and Scarnera, M. and Schaile, P. and Schaufel, M. and Schieler, H. and Schindler, S. and Schlickmann, L. and Schlüter, B. and Schlüter, F. and Schmeisser, N. and Schmidt, T. and Schröder, F. G. and Schumacher, L. and Schwirn, S. and Sclafani, S. and Seckel, D. and Seen, L. and Seikh, M. and Seunarine, S. and Myhr, P. A. Sevle and Shah, R. and Shefali, S. and Shimizu, N. and Skrzypek, B. and Snihur, R. and Soedingrekso, J. and Sègaard, A. and Soldin, D. and Soldin, P. and Sommani, G. and Spannfellner, C. and Spiczak, G. M. and Spiering, C. and Stachurska, J. and Stamatikos, M. and Stanev, T. and Stezelberger, T. and Stürwald, T. and Stuttard, T. and Sullivan, G. W. and Taboada, I. and Ter-Antonyan, S. and Terliuk, A. and Thakuri, A. and Thiesmeyer, M. and Thompson, W. G. and Thwaites, J. and Tilav, S. and Tollefson, K. and Toscano, S. and Tosi, D. and Trettin, A. and Upadhyay, A. K. and Upshaw, K. and Vaidyanathan, A. and Valtonen-Mattila, N. and Valverde, J. and Vandenbroucke, J. and Van Eeden, T. and Van Eijndhoven, N. and Van Rootselaar, L. and Van Santen, J. and Vara, J. and Varsi, F. and Venugopal, M. and Vereecken, M. and Carrasco, S. Vergara and Verpoest, S. and Veske, D. and Vijai, A. and Villarreal, J. and Walck, C. and Wang, A. and Warrick, E. H. S. and Weaver, C. and Weigel, P. and Weindl, A. and Weldert, J. and Wen, A. Y. and Wendt, C. and Werthebach, J. and Weyrauch, M. and Whitehorn, N. and Wiebusch, C. H. and Williams, D. R. and Witthaus, L. and Wolf, M. and Wrede, G. and Xu, X. W. and Yanez, J. P. and Yao, Y. and Yildizci, E. and Yoshida, S. and Young, R. and Yu, F. and Yu, S. and Yuan, T. and Jurowitzki, A. Zander and Zegarelli, A. and Zhang, S. and Zhang, Z. and Zhelnin, P. and Zilberman, P. and IceCube Collaboration}, year={2026}, month=mar, pages={L26} }

@article{icecube2023observation,
  title={Observation of high-energy neutrinos from the Galactic plane},
  author={Abbasi, R and Ackermann, M and Adams, J and Aguilar, JA and Ahlers, M and Ahrens, M and Alameddine, JM and Alves Jr, AA and Amin, NM and others},
  journal={Science},
  volume={380},
  number={6652},
  pages={1338--1343},
  year={2023},
  publisher={American Association for the Advancement of Science}
}

@article{IceCube_Collaboration_2025, title={Physics potential of the IceCube Upgrade for atmospheric neutrino oscillations}, rights={Creative Commons Attribution 4.0 International}, url={https://arxiv.org/abs/2509.13066}, DOI={10.48550/ARXIV.2509.13066}, publisher={arXiv}, author={Abbasi, R. and Ackermann, M. and Adams, J. and Aguilar, J. A. and Ahlers, M. and Ahrens, M. and Alameddine, J. M. and Alispach, C. and Alves, A. A. and Amin, N. M. and others}, year={2025} }

@article{Braun_Baker_Dumm_Finley_Karle_Montaruli_2010, title={Time-dependent point source search methods in high energy neutrino astronomy}, volume={33}, rights={https://www.elsevier.com/tdm/userlicense/1.0/}, ISSN={09276505}, DOI={10.1016/j.astropartphys.2010.01.005}, number={3}, journal={Astroparticle Physics}, author={Braun, Jim and Baker, Mike and Dumm, Jon and Finley, Chad and Karle, Albrecht and Montaruli, Teresa}, year={2010}, month=apr, pages={175–181}, language={en} }

@article{Wright_Eisenhardt_Mainzer_Ressler_Cutri_Jarrett_Kirkpatrick_Padgett_McMillan_Skrutskie_et_al._2010, title={THE WIDE-FIELD INFRARED SURVEY EXPLORER (WISE): MISSION DESCRIPTION AND INITIAL ON-ORBIT PERFORMANCE}, volume={140}, ISSN={0004-6256, 1538-3881}, DOI={10.1088/0004-6256/140/6/1868}, number={6}, journal={The Astronomical Journal}, author={Wright, Edward L. and Eisenhardt, Peter R. M. and Mainzer, Amy K. and Ressler, Michael E. and Cutri, Roc M. and Jarrett, Thomas and Kirkpatrick, J. Davy and Padgett, Deborah and McMillan, Robert S. and Skrutskie, Michael and Stanford, S. A. and Cohen, Martin and Walker, Russell G. and Mather, John C. and Leisawitz, David and Gautier, Thomas N. and McLean, Ian and Benford, Dominic and Lonsdale, Carol J. and Blain, Andrew and Mendez, Bryan and Irace, William R. and Duval, Valerie and Liu, Fengchuan and Royer, Don and Heinrichsen, Ingolf and Howard, Joan and Shannon, Mark and Kendall, Martha and Walsh, Amy L. and Larsen, Mark and Cardon, Joel G. and Schick, Scott and Schwalm, Mark and Abid, Mohamed and Fabinsky, Beth and Naes, Larry and Tsai, Chao-Wei}, year={2010}, month=dec, pages={1868–1881} }

@article{Mainzer_Bauer_Cutri_Grav_Masiero_Beck_Clarkson_Conrow_Dailey_Eisenhardt_et_al._2014, title={INITIAL PERFORMANCE OF THE NEOWISE REACTIVATION MISSION}, volume={792}, rights={http://iopscience.iop.org/info/page/text-and-data-mining}, ISSN={1538-4357}, DOI={10.1088/0004-637X/792/1/30}, number={1}, journal={The Astrophysical Journal}, author={Mainzer, A. and Bauer, J. and Cutri, R. M. and Grav, T. and Masiero, J. and Beck, R. and Clarkson, P. and Conrow, T. and Dailey, J. and Eisenhardt, P. and Fabinsky, B. and Fajardo-Acosta, S. and Fowler, J. and Gelino, C. and Grillmair, C. and Heinrichsen, I. and Kendall, M. and Kirkpatrick, J. Davy and Liu, F. and Masci, F. and McCallon, H. and Nugent, C. R. and Papin, M. and Rice, E. and Royer, D. and Ryan, T. and Sevilla, P. and Sonnett, S. and Stevenson, R. and Thompson, D. B. and Wheelock, S. and Wiemer, D. and Wittman, M. and Wright, E. and Yan, L.}, year={2014}, month=aug, pages={30} }

\begin{appendix}
\section{Population study}
\label{appendix}
\nolinenumbers

\begin{figure}[h!]
\centering
    \centering
    \includegraphics[width=\linewidth]{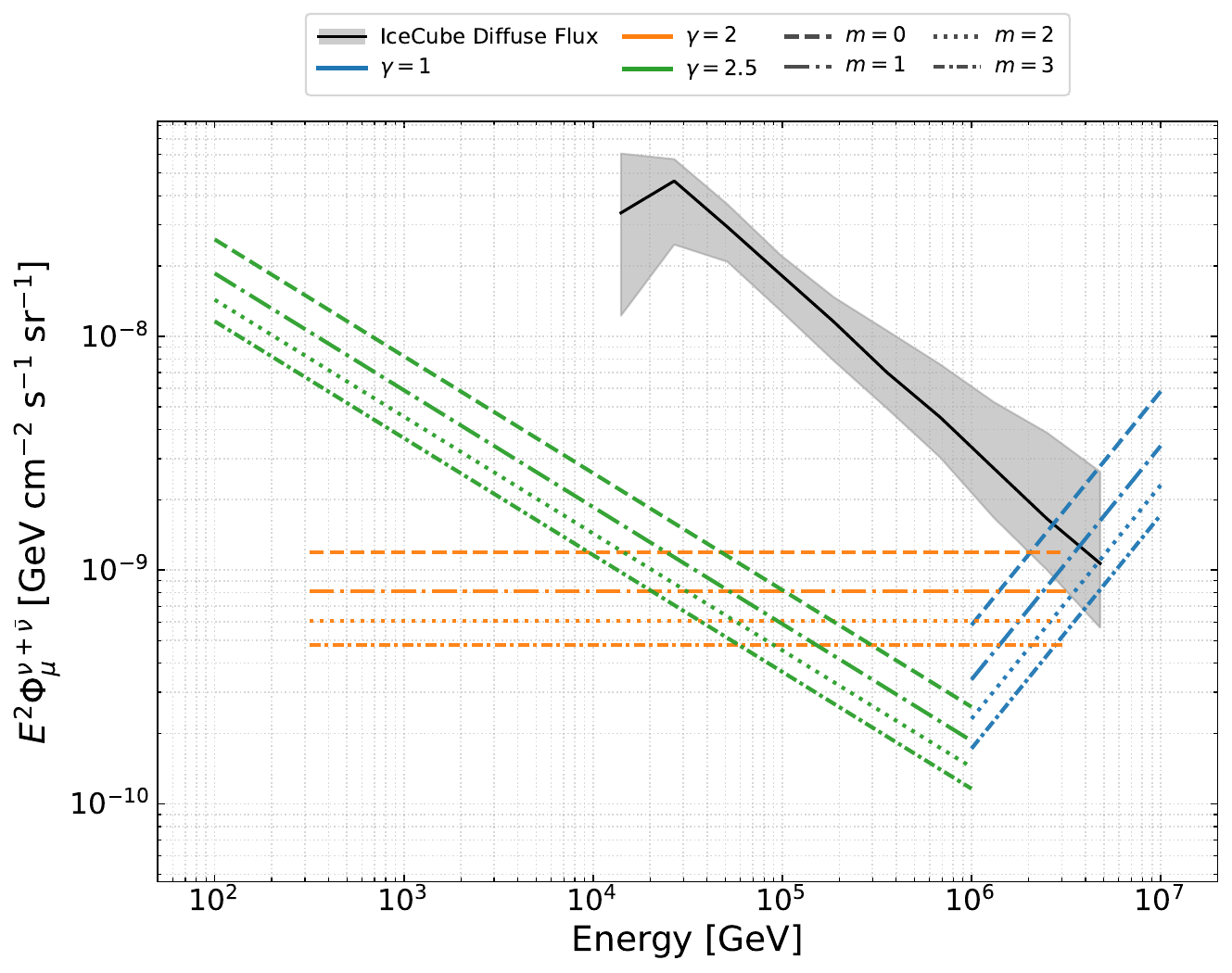}
\caption{Upper limits on the muon-neutrino flux from infrared flares in the Northern Sky are shown for three assumed spectral indices ($\gamma = 1, 2$ and $2.5$). For alternative cosmological source-evolution models, parameterized as $f(z)=(1+z)^{-m}$, where $z$ is the cosmological redshift and $m$ determines the strength of the redshift evolution, different line styles indicate the corresponding values of $m$. The gray band indicates the 95$\%$ confidence interval of the diffuse astrophysical neutrino flux measured by IceCube \citep{Abbasi_Ackermann_Adams_Agarwalla_Aguilar_Ahlers_Alameddine_Ali_Amin_Andeen_et_al._2026_break}. The diffuse flux is shown assuming a single-flavor normalization.}
\label{fig:model_pop_study}
\end{figure}

Since the cosmological evolution of IR-selected TDEs is not yet well constrained observationally, we investigate how the inferred population limits depend on the assumed source evolution. To illustrate this dependence, we show in Fig.~\ref{fig:model_pop_study} the ULs obtained for alternative parameterizations of the source volumetric rate, $f(z)=(1+z)^{-m}$, for different values of $m$. The parameter m controls how rapidly the volumetric source rate decreases with increasing redshift. Larger values of m decrease the contribution from distant sources, reducing the cumulative diffuse neutrino flux predicted for a given source luminosity. Consequently, the inferred diffuse-flux ULs become more stringent. Across the range of models considered here, the resulting ULs vary by up to a factor of $\sim 2.5$ relative to the baseline evolution adopted from \citet{sun2015extragalactic}.

\end{appendix}

\end{document}